%% file: paperQDJJsub2.tex
\documentclass[english,aps,prb,superscriptaddress,groupaddress,reprint,longbibliography]{revtex4-2}
\usepackage[T1]{fontenc}
\usepackage[latin9]{inputenc}
\usepackage[pdftex]{color}
\usepackage{bm}
\usepackage{amsmath}
\usepackage{amssymb}
\usepackage[pdftex]{graphicx}

\makeatletter

\usepackage{braket} %\Braket, \Bra, \Ket
\newcommand{\up}{\uparrow}
\newcommand{\dw}{\downarrow}

\usepackage[svgnames]{xcolor}

\usepackage{hyperref}
\hypersetup{
    colorlinks=true,
    linkcolor=blue,
    filecolor=magenta,      
    urlcolor=magenta,
    citecolor=violet}

\usepackage[capitalise]{cleveref}
\AtBeginDocument{%
\let\ref\cref
}
\usepackage{prettyref}
\newrefformat{rule}{Rule \labelcref{#1}}

\makeatother

\usepackage{babel}
\begin{document}
\title{Non-equilibrium cotunneling in quantum dot Josephson junctions }
\author{Jordi Pic\'{o}-Cort\'{e}s}
\affiliation{Institute for Theoretical Physics, University of Regensburg, 93040
Regensburg, Germany.}
\affiliation{Instituto de Ciencia de Materiales de Madrid (CSIC) 28049 Madrid,
Spain.}
\author{Gloria Platero}
\affiliation{Instituto de Ciencia de Materiales de Madrid (CSIC) 28049 Madrid,
Spain.}
\author{Andrea Donarini}
\affiliation{Institute for Theoretical Physics, University of Regensburg, 93040
Regensburg, Germany.}
\author{Milena Grifoni}
\affiliation{Institute for Theoretical Physics, University of Regensburg, 93040
Regensburg, Germany.}
\begin{abstract}
We investigate non-equilibrium transport through superconducting nanojunctions
using a Liouville space approach. The formalism allows us to study
finite gap effects, and to account for both quasiparticle and Cooper
pair tunneling. With focus on the weak tunneling limit, we study the
stationary dc and ac current up to second order (cotunneling) in the
hybridization energy. For the particular case of a strongly interacting
quantum dot sandwiched between two superconductors, we identify the
characteristic virtual processes that yield the Andreev and Josephson
current and obtain the dependence on the gate and bias voltage for
the dc current, the critical current and the phase-dependent dissipative
current. In particular, the critical current is characterized by regions
in the stability diagram in which its sign changes from positive to
negative, resulting in a multitude of $0-\pi$ transitions. The latter
signal the interplay between strong interactions and tunneling at
finite bias. 
\end{abstract}
\maketitle

\section{Introduction}

\global\long\def\mV{\text{mV}}%

\global\long\def\meV{\text{meV}}%

\global\long\def\mueV{\text{\textgreek{m}eV}}%

\global\long\def\real{\text{Re}}%

\global\long\def\imag{\text{Im}}%

\global\long\def\lapt{\mathrm{L}}%

\global\long\def\fsop{\mathcal{D}}%

\global\long\def\dos{G}%

\global\long\def\g{g}%

\global\long\def\ai{\psi}%

Advancements in nanostructure fabrication have allowed the manufacturing
of increasingly smaller Josephson junctions. For mesoscopic and nano-sized
junctions, genuine quantum mechanical effects often rule the transport
properties of the device. Superconducting junctions based on quantum
point contacts were among the first of such systems to be studied
in detail~\citep{Cuevas1996,Cuevas2002,Golubov2004}. For short weak
links with few channels, multiple coherent Andreev reflections give
rise to current-carrying Andreev bound states which can be employed
to effectively characterize the properties of the Josephson current~\citep{Bretheau2013}.
These ideas have been used extensively in recent years to investigate
novel superconducting transport phenomena, such as the fractional
Josephson effect in topological junctions~\citep{Oreg2010,Cook2011,Stanescu2013},
non-reciprocal transport~\citep{Tokura2018,Baumgartner2021,Baumgartner2022},
and multi-terminal junctions~\citep{Pfeffer2014,Meyer2017}. 

As the size of the junction is decreased even further, Coulomb interaction
effects between the strongly confined charge carriers becomes more
important. Josephson junctions based on quantum dots (QDs), have been
studied in detail both experimentally~\citep{Doh2005,Cleuziou2006,Pillet2010,Dirks2011,DeLagrange2015}
and theoretically~\citep{Siano2004,Sellier2005,Karrasch2008}. They
have further received attention as a platform for quantum computation,
in the form of Andreev spin qubits~\citep{Chtchelkatchev2003,Hays2021}.
The combination of superconducting correlations and electronic interactions
yields a rich phenomenology~\citep{MartnRodero2011,Meden2019}. A
hallmark of this interplay is the $0-\pi$ transition, occurring in
QDs with odd occupancy. Here the sign of the equilibrium critical
current changes as the junction's ground state turns from a singlet
to a doublet~\citep{vanDam2006,Jorgensen2007,Maurand2012,Delagrange2016,Li2017}.
For strongly coupled QDs, Kondo correlations further affect the doublet
and singlet phases~\citep{Eichler2009,Deacon2010,Pillet2013,Bargerbos2022}.
This work extends these results to a non-equilibrium situation. In
particular, we show that a multitude of $0-\pi$ transitions can be
observed in weakly coupled junctions, in a delicate interplay between
interaction effects and Cooper pair tunneling at finite bias.

The Green's function formalism is the method of choice for non-interacting
junctions. While it has been extended to interacting systems via mean
field~\citep{Rozhkov1999,MartnRodero2012} and perturbative expansions~\citep{Clerk2000,Vecino2003},
the regime of interest for Josephson junctions often involves interaction
strengths that lie beyond the applicability of these approximation
schemes~\citep{DeFranceschi2010}. Alternatively, the tunnel coupling
to the leads can be treated perturbatively while keeping the interaction
exactly~\citep{Glazman1989,Choi2000,Novotny2005}. In order to perform
this expansion systematically, a generalized master equation (GME)
for the density operator can be employed~\citep{Schoeller1994,Konig1996,Konig1996b},
in which the distinct tunneling processes can be identified using
diagrammatic techniques. This formalism can furthermore be extended
to larger tunneling amplitudes via diagrammatic resummations~\citep{Kern2013,Magazzu2022,GrifoniBook}
and renormalization group methods~\citep{Saptsov2012}. 

The GME approach has found applications in the study of QD-based Josephson
junctions. This includes three-terminal junctions with one non superconducting
lead which acts as a tunneling probe~\citep{Pala2007,Governale2008,Moghaddam2012},
Cooper pair splitters~\citep{Recher2001,Probst2016,Hussein2016,Hussein2019}
and Andreev bound state spectra~\citep{Pillet2010,Kumar2014,Kirsanskas2015}.
Many of these works employ other approximations on top of weak coupling,
such as the assumption of infinite gap or infinite interaction. The
former, in particular, allows for resummation techniques that restore
the Andreev bound state picture~\citep{Eldridge2010,Sothmann2014,Sothmann2015},
and can even be extended to capture perturbatively the effect of the
quasiparticles~\citep{Futterer2013}. Recent works have employed
fermionic duality to investigate the dynamics in the time domain of
junctions with large gap superconductors~\citep{Ortmanns2023,Ortmanns2023b}. 

Non-equilibrium properties of Josephson junctions through interacting
quantum dots  have received a comparatively poor attention so far.
This regime offers, nonetheless, a rich phenomenology~\citep{DellAnna2008,Jonckheere2009,Andersen2011,Hiltscher2012,Rentrop2014,Lamic2020}.
The finite bias properties of a carbon-nanotube based Josephson QD
have been probed in a recent experiment by the measurement of the
Josephson radiation~\citep{Watfa2021} with intriguing unexplained
features. This is the topic of the present work. 

In the non-equilibrium case the GME approach is often used in a particle
conserving framework of superconductivity~\citep{Josephson1962},
since it naturally allows to account for finite bias voltages across
the junction~\citep{Pala2007,Hiltscher2012}. In a recent publication~\citep{Siegl2023},
some of us have employed the GME approach to investigate the interplay
of a dc and an ac bias voltage on the transport characteristics of
a Josephson quantum dot at finite gaps, temperature and interaction.
In that study it was shown that, already in lowest order in the tunneling
coupling, intriguing features may occur, such as an inversion of the
dc-current due to the combination of the ac drive and a non-flat density
of states. 

Here we focus solely on the effects of a dc voltage bias on transport
and fully include the first two non-vanishing orders in the tunnel
coupling.  We provide comprehensive numerical and analytical results
for both the dc component of the current as well as its first harmonic
in the ac Josephson frequency. For the latter, which provides the
Josephson current out of equilibrium, we find two distinct contributions:
the critical current, which may be non-zero at zero bias, and a phase-dependent
dissipative term, sometimes referred to as the $\cos\varphi$ term~\citep{Harris1974,Waldram1976}.
Knowledge of both allows us to evaluate the Josephson radiation, hence
providing a direct benchmark to present cutting-edge experiments~\citep{Deacon2017,Watfa2021}. 

As we shall see, accounting for the second order is fundamental, as
the Josephson current is of this order for most situations. Inspection
of the stability diagrams allows us to identify the relevant transport
processes, being mediated solely by quasiparticles, Cooper pairs or
a combination of both. For example, subgap features from thermally
excited quasiparticles permeate the dc-component of the current. Cotunneling-like
processes are responsible for a $\pi-$phase of the critical current
in the odd Coulomb diamond at low bias, extending results found for
the zero bias case~\citep{Glazman1989}. 

The manuscript is organized as follows. \ref{sec:Theoretical-framework}
introduces the model and the transport theory that we will employ
throughout this work. In particular, in \ref{subsec:The-Josephson-effect}
we show how the ac Josephson effect appears naturally in the system
as a consequence of considering arbitrary coherences involving Cooper
pairs. In \ref{sec:Perturbation-Theory} we introduce the perturbative
scheme employed to calculate the steady state density operator and
the current. In \ref{sec:dc-current} and \ref{sec:Josephson-current}
we present numerical results for the dc and ac components of the current
and highlight their main features. Finally, in \ref{sec:Discussion}
we summarize our results and provide an outlook for future research.

\section{Theoretical framework\label{sec:Theoretical-framework}}

\subsection{Model Hamiltonian\label{subsec:Theoretical-model}}

We focus on the archetypal model of an interacting Josephson junction:
the single impurity Anderson model (SIAM) with superconducting leads.
The total \hbox{Hamiltonian} $\hat{H}=\hat{H}_{SC}+\hat{H}_{QD}+\hat{H}_{T}$
includes the contributions from the leads $\hat{H}_{SC}$, the quantum
dot $\hat{H}_{QD}$, and the tunnel coupling between them $\hat{H}_{T}$.
The last two are given respectively by
\begin{align}
\hat{H}_{QD} & =\sum_{\sigma}\epsilon_{d}\hat{d}_{\sigma}^{\dagger}\hat{d}_{\sigma}+U\hat{d}_{\up}^{\dagger}\hat{d}_{\up}\hat{d}_{\dw}^{\dagger}\hat{d}_{\dw},\label{eq:SIAM-H}\\
\hat{H}_{T} & =\sum_{l\bm{k}\sigma}\left(t_{l\bm{k}}\hat{c}_{l\bm{k}\sigma}^{\dagger}\hat{d}_{\sigma}+t_{l\bm{k}}^{*}\hat{d}_{\sigma}^{\dagger}\hat{c}_{l\bm{k}\sigma}\right).\label{eq:HT}
\end{align}
Here, $\hat{d}_{\sigma}^{(\dagger)}$ are annihilation (creation)
operators of the dot with spin $\sigma=\,\up,\dw$, while $\epsilon_{d}$
is the single particle energy of the dot, tunable via an electric
gate with voltage $V_{g}$. We choose in the following $\epsilon_{d}=-eV_{g}$.
Finally, $U$ is the Coulomb interaction. The tunneling process is
characterized by spin conserving tunneling amplitudes $t_{l\bm{k}}$
for the transfer of a dot electron to lead $l=L,R$, where it is described
by annihilation (creation) fermionic operators $\hat{c}_{l\bm{k}\sigma}^{(\dagger)}$
with momentum $\bm{k}$ and spin $\sigma$. 

In order to account for charge transfer out of equilibrium, we treat
the superconductors in a particle-conserving formulation~\citep{LeggettBook,PicoCortesTh2021}.
In this framework, the bare fermion operators can be written, employing
the particle-conserving Bogoliubov-Valatin transformations, as

\begin{equation}
\hat{c}_{l\bm{k}\sigma}^{\dagger}=u_{l\bm{k}}\hat{\gamma}_{\bm{k}\sigma}^{\dagger}+\sigma v_{l\bm{k}}^{*}\hat{\gamma}_{l\bar{\bm{k}}\bar{\sigma}}\hat{S}_{l}^{\dagger},\label{eq:Bog-Valatin}
\end{equation}
where $\sigma=\pm$ for $\up/\dw$, respectively, and 
\begin{align}
u_{l\bm{k}} & =\sqrt{(1/2)\left(1+\xi_{l\bm{k}}/E_{l\bm{k}}\right)},\\
v_{l\bm{k}} & =e^{i\phi_{l}}\sqrt{\left(1/2\right)\left(1-\xi_{l\bm{k}}/E_{l\bm{k}}\right)},
\end{align}
with $E_{l\bm{k}}=\sqrt{\xi_{l\bm{k}}^{2}+|\Delta_{l}|^{2}}$ the
quasiparticle excitation energy for mode $\bm{k}$. Here, $\xi_{l\bm{k}}$
describes the single particle dispersion of the bare fermions measured
from the chemical potential $\mu_{l}$, $\Delta_{l}$ is the superconducting
order parameter, and $\phi_{l}=\text{arg}\left(\Delta_{l}\right)$.
We take $\phi_{l}=0$ in this work without loss of generality. More
importantly, we have employed the Bogoliubov operators $\hat{\gamma}_{l\bm{k}\sigma}^{\dagger},\hat{\gamma}_{l\bm{k}\sigma}$
which describe quasiparticle excitations, as well as Cooper pair operators
$\hat{S}_{l},\hat{S}_{l}^{\dagger}$ which destroy or create a Cooper
pair in the condensate~\citep{Josephson1962}. Hence, the Bogoliubov-Valatin
transformation conserves the total charge. The first term in \ref{eq:Bog-Valatin}
describes an electron-like excitation, while the second is associated
with a hole-like excitation~\citep{TinkhamBook}. 

The quasiparticle states are characterized by the occupation $\nu_{l\bm{k}\sigma}=0,1$
of the fermionic modes, and the state of the Cooper pair condensate
by the total number of Cooper pairs, which we denote by $M_{l}$.
With this, we can write an arbitrary state of lead $l$ as $\ket{M_{l}}\ket{\{\nu_{l\bm{k}\sigma}\}}$,
with total number of electrons (Cooper pairs and quasiparticles) equal
to $N_{l}=2M_{l}+\sum_{\bm{k}\sigma}\nu_{l\bm{k}\sigma}$. The effect
of the Cooper pair operators on any such state is given by
\begin{equation}
\hat{S}_{l}^{\dagger}\ket{M_{l}}\ket{\{\nu_{l\bm{k}\sigma}\}}=\ket{M_{l}+1}\ket{\{\nu_{l\bm{k}\sigma}\}}.\label{eq:action-S-op}
\end{equation}
Similarly, $\hat{S}_{l}$ reduces $M_{l}$ by 1 and leaves the occupation
of the quasiparticle states invariant. Employing this, we see that
the Cooper pair and quasiparticle operators commute~\citep{Pfaller2013}
\begin{align}
[\hat{S}_{l}^{\dagger},\hat{\gamma}_{l\bm{k}\sigma}^{\dagger}] & =[\hat{S}_{l}^{\dagger},\hat{\gamma}_{l\bm{k}\sigma}]=0.\label{eq:comm-rels}
\end{align}
Moreover, the Cooper pair creation and annihilation operators satisfy
\begin{equation}
\hat{S}_{l}\hat{S}_{l}^{\dagger}=1,\,\,\,[\hat{S}_{l},\hat{S}_{l}^{\dagger}]\propto\mathcal{O}\left(1/M_{l}\right).
\end{equation}
Assuming that the Cooper pair operators commute is a good approximation
in the limit of large Cooper pair numbers and we do so from here onward.

Upon employing \ref{eq:Bog-Valatin}, we can divide the Hamiltonian
for the leads into two contributions, $\hat{H}_{SC}=\hat{H}_{QP}+\hat{H}_{CP}.$
The former describes the quasiparticle excitations and reads

\begin{align}
\hat{H}_{QP} & =\sum_{l\bm{k}\sigma}E_{l\bm{k}}\hat{\gamma}_{l\bm{k}\sigma}^{\dagger}\hat{\gamma}_{l\bm{k}\sigma}+\sum_{l}\mu_{l}\hat{N}_{QP,l},\label{eq:HQP}\\
\hat{N}_{QP,l} & =\sum_{\bm{k}\sigma}\hat{\gamma}_{l\bm{k}\sigma}^{\dagger}\hat{\gamma}_{l\bm{k}\sigma},
\end{align}
where $\mu_{l}=a_{l}eV_{b}$ is the chemical potential of lead $l$,
$V_{b}$ is the bias voltage and the coefficients $a_{l}$ satisfy
$a_{L}-a_{R}=1$. The latter contribution $\hat{H}_{CP}$ describes
the Cooper pair condensate and has the form
\begin{align}
\hat{H}_{CP} & =\sum_{l}\mu_{l}\hat{N}_{CP,l},\quad\hat{N}_{CP,l}\ket{M_{l}}=2M_{l}\ket{M_{l}}.
\end{align}
Together with \ref{eq:action-S-op} and the corresponding equation
for $\hat{S}_{l}$, it follows that $[\hat{S}_{l},\hat{H}_{CP}]=2\mu_{l}\hat{S}_{l}.$ 

The Hamiltonian $\hat{H}_{SC}$ corresponds to a mean field approximation
to the true Hamiltonian of the superconductor. By definition, the
quasiparticles operators describe excitations on top of the ground
state. Within the mean field approximation, the quasiparticle spectrum
is assumed to be independent of the number of Cooper pairs in the
ground state, yielding the commutation relations in \ref{eq:comm-rels}.
As a result of these relations, we find $[\hat{S}_{l},\hat{H}_{T}]\sim[\hat{S}_{l},\hat{S}_{l}^{\dagger}]\sim0$
in the approximation of large Cooper pair numbers, but $[\hat{H}_{T},\hat{N}_{CP}]\neq0$.
The consequences and validity of the approximation will be further
discussed in \ref{subsec:The-Josephson-effect}. We anticipate that,
since the quasiparticles are the only source of dissipation in the
model, the dynamics of the Cooper pair operators are completely coherent
and their time evolution is fully determined by the Cooper pair Hamiltonian.

\subsection{Transport theory\label{subsec:Transport-theory}}

Interacting open quantum systems can be studied via a many-body generalized
master equation (GME) based on the Nakajima-Zwanzig formalism~\citep{Nakajima1958,Zwanzig1960}.
The GME describes the dynamics of a central system (denoted by $S$
from here on) under the effect of the coupling to a bath (denoted
by $B$). In the context of transport, the system is often considered
to be the nanostructure through which the current flows (in this case,
the quantum dot) and the bath the electrodes to which it is coupled.
However, we note that the distinction between ``system'' and ``bath''
is, in the Nakajima-Zwanzig formalism, purely one of convenience.
Indeed, we will make use of this fact shortly in order to account
for the peculiarities of the superconducting case.

We define the reduced density operator as $\hat{\rho}_{S}\left(t\right)=\text{Tr}_{B}\left\{ \hat{\rho}\left(t\right)\right\} $,
where $\text{Tr}_{k}\left\{ \cdots\right\} $ indicates partial trace
over subsystem $k$, and $\hat{\rho}\left(t\right)$ is the density
operator of the coupled bath and system; it satisfies the Liouville-von
Neumann equation. The reduced density operator in turn obeys the GME~\citep{Schoeller1994,Konig1996,Konig1996b}
\begin{equation}
\frac{d}{dt}\hat{\rho}_{S}\left(t\right)=\mathcal{L}_{S}\hat{\rho}_{S}\left(t\right)+\int_{0}^{t}ds\,\mathcal{K}_{S}\left(t-s\right)\hat{\rho}_{S}\left(s\right),\label{eq:Nakajima-Zwanzig-eq}
\end{equation}
where $i\hbar\mathcal{L}_{k}\hat{O}=[\hat{H}_{k},\hat{O}]$ is the
Liouvillian superoperator associated with the Hamiltonian $\hat{H}_{k}$
and $\mathcal{K}_{S}\left(t\right)$ is the propagation kernel. The
first term on the right hand side fully encodes the coherent dynamics
of the isolated system via $\mathcal{L}_{S}$, and the second the
effect of the bath and its coupling to the system exactly via $\mathcal{K}_{S}\left(t\right)$. 

Once the reduced density operator is known, the current through the
junction can be calculated as the statistical average of the current
operator. For lead $l$, this is given by $\hat{I}_{l,H}\left(t\right)=-e\,d\hat{N}_{l,H}\left(t\right)/dt$,
with $\hat{O}_{l,H}\left(t\right)$ referring to the Heisenberg picture
representation of operator $\hat{O}_{l}$. Notice that $\hat{N}_{l}$
accounts for both Cooper pairs and quasiparticles. Here and onward
we consider the symmetrized current $\hat{I}=(1/2)(\hat{I}_{L}-\hat{I}_{R})$.
This expression avoids the issue of displacement currents~\citep{Bruder1994,Hiltscher2012}
which appear in time-dependent problems. Then, the statistical average
of the current operator can be calculated as 

\begin{align}
I\left(t\right) & =\langle\hat{I}\left(t\right)\rangle=\text{Tr}_{S}\left\{ \int_{0}^{t}ds\,\mathcal{J}_{S}\left(t-s\right)\hat{\rho}_{S}\left(s\right)\right\} ,\label{eq:current-red}
\end{align}
where $\mathcal{J}_{S}\left(t-s\right)$ is the current kernel. Both
$\mathcal{K}_{S}\left(t-s\right)$ and $\mathcal{J}_{S}\left(t-s\right)$
are superoperators acting on the system degrees of freedom only. Their
expressions are given in \ref{sec:NZE-formalism}.

\subsection{The stationary state}

In the following, we focus on the stationary dynamics of the problem
at $t\to\infty$. We start by considering solutions of \ref{eq:Nakajima-Zwanzig-eq}
of the form
\begin{align}
\hat{\rho}_{S}\left(t\right) & =\sum_{\lambda_{i}}e^{\lambda_{i}t}\hat{r}_{S}\left(\lambda_{i}\right),\label{eq:sol-NZE}\\
\hat{r}_{S}\left(\lambda_{i}\right) & =\lim_{\lambda\to\lambda_{i}}\left(\lambda-\lambda_{i}\right)\lapt\{\hat{\rho}_{S}\}(\lambda),\label{eq:residues-lambda}
\end{align}
where $\lapt\{f\}\left(\lambda\right)=\int_{0}^{\infty}dte^{-\lambda t}f\left(t\right)$
is the Laplace transform of $f\left(t\right)$. The residue operators
$\hat{r}_{S}\left(\lambda_{i}\right)$ with $\real\{\lambda_{i}\}>0$
in \ref{eq:sol-NZE} are unphysical and those with $\real\{\lambda_{i}\}<0$
are transients which vanish in the stationary state. As such, we will
only be concerned with the solutions satisfying $\real\{\lambda_{i}\}=0$.
Hermiticity of the density operator requires that for each $\lambda_{i}$
its complex conjugate $\lambda_{i}^{*}$ must also appear in \ref{eq:sol-NZE},
with $\hat{r}_{S}\left(\lambda_{i}^{*}\right)=\hat{r}_{S}^{\dagger}\left(\lambda_{i}\right)$.
Furthermore, $\lambda_{i}=0$ must always be a solution in order to
satisfy conservation of probability at all times. By the same reason
it follows that $\text{Tr}_{S}\{\hat{r}_{S}\left(\lambda_{i}\right)\}=\delta_{\lambda_{i},0}$. 

The $\hat{r}_{S}\left(\lambda_{i}\right)$ in \ref{eq:sol-NZE} are
the solutions of the equation
\begin{equation}
\bigl[\mathcal{L}_{S}-\lambda+\tilde{\mathcal{K}}_{S}(\lambda)\bigr]\hat{r}_{S}\left(\lambda\right)=0.\label{eq:Liouville-transf-NZE}
\end{equation}
Here, we use the short-hand notation $\tilde{\mathcal{K}}_{S}(\lambda)\equiv\lapt\{\mathcal{K}_{S}\}\left(\lambda\right)$
for the Laplace transform of $\mathcal{K}_{S}\left(t\right)$, given
by (see~\ref{sec:NZE-formalism})
\begin{align}
 & \tilde{\mathcal{K}}_{S}(\lambda)\bullet=\label{eq:reduced-kernel-2}\\
 & \text{Tr}_{B}\biggl\{\mathcal{L}_{T}\sum_{k=0}^{\infty}\bigl(\tilde{\mathcal{G}_{0}}(\lambda)\mathcal{Q}\mathcal{L}_{T}\mathcal{Q}\bigr)^{2k}\tilde{\mathcal{G}_{0}}(\lambda)\mathcal{L}_{T}\bullet\otimes\hat{\rho}_{B}\biggr\},\nonumber 
\end{align}
where we introduced the projector $\mathcal{Q}\bullet=\bullet-\text{Tr}_{B}\{\bullet\}\otimes\hat{\rho}_{B}$,
with $\hat{\rho}_{B}$ a reference density operator of the bath that
satisfies $\mathcal{L}_{B}\hat{\rho}_{B}=0$, and the Laplace-transformed
free propagator 
\begin{equation}
\tilde{\mathcal{G}_{0}}(\lambda)=\frac{1}{\lambda-\mathcal{L}_{S}-\mathcal{L}_{B}}.\label{eq:propagator}
\end{equation}
The superoperator $\tilde{\mathcal{K}}_{S}(\lambda)$ exists provided
that $\mathcal{K}_{S}\left(t\right)$ is bounded for $t>0$, which
we assume in the following. We remark that only even factors of the
tunneling amplitude appear in \ref{eq:reduced-kernel-2} due to conservation
of particle number and the overall trace over the quasiparticle bath. 

The exponents $\lambda_{i}$ appearing in \ref{eq:sol-NZE} are the
solutions of a non-linear eigenvalue equation, which can be hard to
find even for simple systems. In the case of a Josephson quantum
dot, symmetry arguments allow one to obtain exactly the subset of
eigenvalues $\{\lambda_{i}\}$ governing the stationary dynamics,
as we will show in \ref{subsec:The-Josephson-effect}.

The current for a solution of the form of \ref{eq:Liouville-transf-NZE}
can be obtained via 
\begin{equation}
I\left(t\right)=\sum_{\lambda_{i}}\text{Tr}_{S}\biggl\{\tilde{\mathcal{J}}_{S}(\lambda_{i})\hat{r}_{S}\left(\lambda_{i}\right)\biggr\} e^{\lambda_{i}t},\label{eq:current-Laplace-transformed-1}
\end{equation}
with $\tilde{\mathcal{J}}_{S}(\lambda)$ the Laplace-transformed current
kernel, resulting from substituting the leftmost instance of $\mathcal{L}_{T}$
for the current operator $\hat{I}$ in \ref{eq:reduced-kernel-2}. 

\subsection{Transport equations for a Josephson junction\label{subsec:The-Cooper-pair}}

We now turn to the question of dealing with the Cooper pair degrees
of freedom within the transport theory outlined above. Within the
mean field description introduced in \ref{subsec:Theoretical-model},
the quasiparticles are fermionic excitations of the superconductor
on top of a Cooper pair condensate. It is thus natural to view the
quasiparticles as the environment and the Cooper pairs as part of
the system~\citep{Pala2007}. We therefore take in the following
\begin{equation}
\hat{H}_{S}=\hat{H}_{QD}+\hat{H}_{CP},\,\,\,\,\,\hat{H}_{B}=\hat{H}_{QP},
\end{equation}
with the reference density operator $\hat{\rho}_{B}$ {[}entering
in \ref{eq:reduced-kernel-2}{]} chosen to be the equilibrium density
operator of the quasiparticles only
\begin{align}
\hat{\rho}_{B}=\frac{1}{Z_{QP}} & e^{-\beta(\hat{H}_{QP}-\sum_{l}\mu_{l}\hat{N}_{QP,l})}.\label{eq:rho-QP}
\end{align}
This choice immediately satisfies $\mathcal{L}_{B}\hat{\rho}_{B}=0$.
Here, $\beta=1/k_{B}T$, with $T$ the temperature of the superconductors,
which we assume to be equal. 

Let us now discuss the consequences of considering the Cooper pairs
as part of the system. To begin with, the degrees of freedom for the
system are now given by $\left(\chi,\bm{M}\right)$~\citep{Governale2008},
where $\chi\in\left\{ 0,\up,\dw,2\right\} $ labels the quantum dot
Fock states $\{\ket{\chi}\}$, and $\bm{M}=\left(M_{L},M_{R}\right)$
is a vector whose elements are the Cooper pair numbers of each superconductor.
As a result, we can write the residue operators as~\citep{Hiltscher2012,Siegl2023}

\begin{equation}
\hat{r}_{S}\left(\lambda\right)=\sum_{\bm{M},\bm{m}}\hat{r}\left(\bm{m},\bm{M};\lambda\right)\ket{\bm{M}+\bm{m}}\bra{\bm{M}},\quad\lambda\in\{\lambda_{i}\},\label{eq:rho-m-M}
\end{equation}
where each $\hat{r}\left(\bm{m},\bm{M};\lambda\right)$ is now an
operator in the QD sector only. Here, $\bm{m}$ describes an imbalance
in Cooper pair numbers between the bra and ket parts of the residue
operator and its components can in principle take any value, positive
or negative. 

Let us note next that 
\begin{align}
i\hbar\mathcal{L}_{CP}\ket{\bm{M}+\bm{m}}\bra{\bm{M}}= & 2\bm{m}\cdot\bm{\mu}\ket{\bm{M}+\bm{m}}\bra{\bm{M}},\label{eq:LCP-effect-NZE-1}
\end{align}
where we have defined a chemical potential vector \hbox{$\bm{\mu}=\left(\mu_{L},\mu_{R}\right)$}.
Due to \ref{eq:LCP-effect-NZE-1}, the kernels are not explicitly
dependent on $\bm{M}$ as only $\mathcal{L}_{CP}$ appears in the
propagators. In fact, only the Cooper pair  ``imbalance'' $\bm{m}$
is actually relevant for the the calculation of the current~\citep{PicoCortesTh2021,Siegl2023}.
In order to take advantage of this, let us define 
\begin{equation}
\hat{r}\left(\bm{m};\lambda\right)=\sum_{\bm{M}}\hat{r}\left(\bm{m},\bm{M};\lambda\right).\label{eq:generalized-partial-tr}
\end{equation}
In particular, $\hat{r}\left(\bm{0};\lambda\right)$ corresponds to
the partial trace over the Cooper pair sector of $\hat{r}_{S}\left(\lambda\right)$.
Meanwhile, the $\hat{r}\left(\bm{m}\neq\bm{0};\lambda\right)$ are
sums over off-diagonals in the Cooper pair sector of the full operator.
Hermiticity of the system density operator requires that $\hat{r}\left(\bm{m};\lambda\right)=[\hat{r}\left(-\bm{m};\lambda^{*}\right)]^{\dagger}$.
Moreover, for any matrix element $[\hat{r}\left(\bm{m}\right)]_{\chi'}^{\chi}=\bra{\chi'}\hat{r}\left(\bm{m}\right)\ket{\chi}$,
the particle number selection rules enforce 
\begin{equation}
N_{\chi'}-N_{\chi}+2\sum_{l}m_{l}=0,\label{eq:particle-cons-coherences}
\end{equation}
with $N_{\chi}$ the number of particles in state $\ket{\chi}$. The
number of particles in the SIAM is limited by the exclusion principle
to be at most $2$. This, together with \ref{eq:particle-cons-coherences},
means that 
\begin{equation}
-1\leq m_{L}+m_{R}\leq1.\label{eq:particle-cons-2}
\end{equation}
However, there is still an infinite number of possible Cooper pair
imbalances $\bm{m}$ because $m_{L}$ and $m_{R}$ are \emph{differences
}in Cooper pair numbers and can be negative. The QD density matrix
elements $[\hat{r}\left(\bm{m}\right)]_{\chi'}^{\chi}$ with $m_{L}+m_{R}=\pm1$
necessarily involve coherences between the QD states with even parity
(i.e. $\ket{0}\bra{2}$ or $\ket{2}\bra{0}$). Meanwhile, terms with
$m_{R}+m_{L}=0$ entail matrix elements of $\hat{r}\left(\bm{m}\right)$
consistent with particle conservation (i.e. the populations and the
coherences $\ket{\up}\bra{\dw},\ket{\dw}\bra{\up}$). Spin coherences
can be neglected in the stationary state due to spin superselection
rules~\citep{GrifoniBook}.

The $\hat{r}\left(\bm{m}\right)$ can be found by directly solving
the set of coupled equations~\citep{Hiltscher2012,PicoCortesTh2021,Siegl2023}
\begin{align}
\left(\mathcal{L}_{QD}-i\omega_{\bm{m}}-\lambda\right)\hat{r}\left(\bm{m};\lambda\right)\nonumber \\
+\sum_{\bm{m}'}\tilde{\mathcal{K}}\left(\bm{m}-\bm{m}';\lambda+i\omega_{\bm{m}'}\right)\hat{r}\left(\bm{m}';\lambda\right) & =0,\label{eq:GME-generalized-partial-tr}
\end{align}
where
\begin{equation}
\omega_{\bm{n}}\equiv2\bm{n}\cdot\bm{\mu}/\hbar,\label{eq:omega_n}
\end{equation}
and
\begin{align}
 & \tilde{\mathcal{K}}\left(\bm{m}-\bm{m}';\lambda+i\omega_{\bm{m}'}\right)=\nonumber \\
 & \sum_{\bm{M}'}\bra{\bm{M}+\bm{m}}[\tilde{\mathcal{K}}_{S}(\lambda)\ket{\bm{M}'+\bm{m}'}\bra{\bm{M}'}]\ket{\bm{M}},\label{eq:kernel-elements}
\end{align}
is a superoperator in the QD space. It can be obtained by collecting
the kernel elements which change the Cooper pair numbers by $\bm{m}-\bm{m}'$.
The shift $\lambda\to\lambda+i\omega_{\bm{m}'}$ in \ref{eq:GME-generalized-partial-tr}
accounts for the action of $\mathcal{L}_{CP}$ after the CP operators
in the kernel have been applied on an initial operator of the form
$\ket{\bm{M}'+\bm{m}'}\bra{\bm{M}'}$, as $\lambda$ and $\mathcal{L}_{CP}$
appear together in the propagators {[}see \ref{eq:propagator}{]}.
The derivation of these expressions is given in more detail in \ref{sec:NZE-sc}.
In particular, see \ref{eq:GME-very-expanded,eq:connection-definitions-kernel}. 

\subsection{The ac Josephson effect\label{subsec:The-Josephson-effect}}

Given the equations obeyed by the residue operators, we can provide
the explicit form of the frequencies $\lambda_{i}$ that are relevant
in the steady state. We first notice that \ref{eq:GME-generalized-partial-tr}
is invariant under the transformation
\begin{equation}
\lambda\to\lambda+i\omega_{\bm{n}},\quad\bm{m}\to\bm{m}-\bm{n}.\label{eq:symmetry-GME}
\end{equation}
Let us consider $\lambda=0$ first. We know that this is always a
solution of the GME, and as such $\hat{r}\left(\bm{m};0\right)$ must
satisfy \ref{eq:GME-generalized-partial-tr}. Then, it follows from
\ref{eq:symmetry-GME} that $\hat{r}\left(\bm{m}-\bm{n};i\omega_{\bm{n}}\right)$
obeys the same equation. In particular, provided that there is a single
solution with $\lambda=0$ to \ref{eq:GME-generalized-partial-tr}
(which we assume in the following), we find
\begin{equation}
\hat{r}\left(\bm{m};i\omega_{\bm{n}}\right)=c_{\bm{n}}\hat{r}\left(\bm{m}+\bm{n};0\right),\label{eq:symmetry-rho}
\end{equation}
for some constant $c_{\bm{n}}$. Therefore, $i\omega_{\bm{n}}$ must
also be valid solutions of \ref{eq:Liouville-transf-NZE} for $\bm{n}\in\mathbb{Z}^{2}$.
As a result, the stationary operator in \ref{eq:sol-NZE} is in general
time-dependent, of the form
\begin{equation}
\hat{\rho}_{S}^{\infty}\left(t\right)=\sum_{\bm{n}}\hat{r}_{S}\left(i\omega_{\bm{n}}\right)e^{i\omega_{\bm{n}}t}.\label{eq:steady-state}
\end{equation}
The coefficients $c_{\bm{n}}$ can be written as 
\begin{align}
c_{\bm{n}} & =\lim_{t\to\infty}C_{\bm{n}}\left(t\right)\nonumber \\
 & =\lim_{t\to\infty}\text{Tr}_{S}\{(\hat{S}_{L}^{\dagger})^{n_{L}}(\hat{S}_{R}^{\dagger})^{n_{R}}e^{-i2\bm{n}\cdot\bm{\mu}t/\hbar}\hat{\rho}_{S}\left(t\right)\}.\label{eq:cn}
\end{align}
They satisfy  $c_{\bm{n}}=c_{-\bm{n}}^{*}$, and $0\leq|c_{\bm{n}}|\leq1$.
The former follows from hermiticity of the solutions of the GME and
the Cooper pair coherences; the latter from the triangle inequality
together with the fact that the density operator is a trace one, non-negative
operator. Furthermore, as the Hamiltonian respects charge conservation,
for any initial state with fixed particle number, the stationary state
will also respect this property~\citep{GrifoniBook}. As a consequence,
only coefficients $c_{\bm{n}}$ with $n_{L}=-n_{R}$ can be non-zero.
This property, together with the definition of the $i\bm{\omega}_{n}$
{[}\ref{eq:omega_n}{]}, means that the stationary time evolution
is periodic in the Josephson frequency 
\begin{equation}
\Omega_{J}=2eV_{b}/\hbar.\label{eq:Josephson-freq}
\end{equation}
As a result of having a time-dependent stationary state, the ac Josephson
effect follows naturally from \ref{eq:current-Laplace-transformed-1}.
Explicitly, the current is given in the stationary state by 
\begin{equation}
I^{\infty}\left(t\right)=\sum_{\bm{n}}I_{\bm{n}}e^{i\omega_{\bm{n}}t},\label{eq:current-steady-state-1}
\end{equation}
with the amplitudes of the current harmonics being
\begin{align}
I_{\bm{n}}= & c_{\bm{n}}\text{Tr}_{QD}\biggl\{\sum_{\bm{m}}\tilde{\mathcal{J}}\left(\bm{n}-\bm{m};i\omega_{\bm{m}}\right)\hat{r}\left(\bm{m};0\right)\biggr\}.\label{eq:current-steady-state-2}
\end{align}
Because the current can be written fully in terms of the operators
$\hat{r}\left(\bm{m};0\right)$, it suffices to calculate only the
$\lambda=0$ components. Hence, in the following we denote $\hat{r}\left(\bm{m};0\right)$
simply by $\hat{r}\left(\bm{m}\right)$. The other harmonics of the
stationary state $\hat{r}\left(\bm{m};i\omega_{\bm{n}}\right)$ can
be obtained from $\hat{r}\left(\bm{m}\right)$ via \ref{eq:symmetry-rho}. 

For the mean field Hamiltonian considered here, the model lacks a
dissipative or dephasing mechanism in the Cooper pair sector, due
to the uncoupling between the condensate and the quasiparticle degrees
of freedom, as discussed in \ref{subsec:Theoretical-model}. As a
result, the coefficients $C_{\bm{n}}\left(t\right)$ defined in \ref{eq:cn}
are time-independent. Then, the $c_{\bm{n}}$ can be formally obtained
from the initial density operator of the junction, which must be provided.
This is discussed in detail in \ref{sec:NZE-sc}. Nonetheless, in
the presence of a dissipation mechanism for the Cooper pair sector,
the information about the initial state will be lost in the steady
state and the value of the coefficients will be set dynamically by
the relevant relaxation mechanism. 

In the following, we consider as an ansatz coefficients of the form
\begin{equation}
c_{\bm{n}}=\delta_{n_{L},-n_{R}}e^{i(n_{R}-n_{L})\varphi_{0}/2},\label{eq:particle-cons-init-cond}
\end{equation}
for a given phase difference $\varphi_{0}$. This choice is motivated
from a reasonable initial state in \ref{sec:NZE-sc}. Moreover,
it reproduces the standard ``BCS'' result for the functional dependence
of the supercurrent on the phase {[}see \ref{eq:I-de-t} below{]},
allowing us to compare directly to previous results in the literature.
A full description of the Cooper pair dynamics, including phase relaxation,
is beyond the scope of this work. 

\section{Perturbation Theory\label{sec:Perturbation-Theory}}

After describing the generalities of the GME and its solutions, we
discuss a perturbative procedure to evaluate the density operator
and the current. The following approach extends the Liouville space
formulation of quantum dot transport developed recently (see e.g.
\citep{GrifoniBook}) to superconductivity, following similar principles
to Refs.~\citep{Pala2007,Governale2008}. 

The problem at hand is made particularly difficult by the combination
of three factors: (1) the coupling with the leads, especially for
a non zero bias voltage, (2) the Coulomb interaction in the quantum
dot, and (3) the superconducting correlations at finite bias, which
result in an infinite set of coupled equations for the residue operator
{[}namely, \ref{eq:GME-generalized-partial-tr}{]}. In order to properly
account for (1) and (2), we treat the kernel perturbatively in the
tunneling Hamiltonian, while keeping the interaction exactly. If we
do so, (3) is also simplified considerably, since only a finite number
of Cooper pairs can be transferred through the junction at any given
perturbation order. As we will show in \ref{sec:Relevance-of-the},
this means that only few terms will be relevant at any given perturbation
level. 

In this work, we focus on the first two non-vanishing perturbation
orders in the tunneling amplitudes. As mentioned above, only even
powers of $\mathcal{L}_{T}$ enter the full expression of the kernel
in \ref{eq:reduced-kernel-2}. As such, it is convenient to introduce
as a perturbative parameter the quantity
\begin{equation}
\Gamma_{l}=\left(2\pi/\hbar\right)\dos_{0,l}|t_{l}|^{2}.
\end{equation}
Here $\dos_{0,l}$ is the density of states (DOS) of lead $l$ in
the normal state, evaluated at the Fermi energy. In the following
we will consider identical leads and tunneling amplitudes, so that
$\Gamma_{l}\equiv\Gamma,$ $\dos_{0,l}\equiv\dos_{0}$ and $\Delta_{l}\equiv\Delta\in\mathbb{R}$
for $l=L,R$, but we will keep the label $l$ in these quantities
for the sake of generality.

The propagation and current kernels are given, to second-to-lowest
order, by 
\begin{align}
\tilde{\mathcal{K}}_{S}\left(\lambda\right) & =\tilde{\mathcal{K}}_{S}^{\left(1\right)}\left(\lambda\right)+\tilde{\mathcal{K}}_{S}^{\left(2\right)}\left(\lambda\right)+\mathcal{O}\left(\Gamma^{3}\right),\label{eq:K-pert}\\
\tilde{\mathcal{J}}_{S}\left(\lambda\right) & =\tilde{\mathcal{J}}_{S}^{\left(1\right)}\left(\lambda\right)+\tilde{\mathcal{J}}_{S}^{\left(2\right)}\left(\lambda\right)+\mathcal{O}\left(\Gamma^{3}\right),\label{eq:J-pert}
\end{align}
with $\tilde{\mathcal{K}}_{S}^{\left(k\right)},\,\tilde{\mathcal{J}}_{S}^{\left(k\right)}\propto\Gamma^{k}$.
These correspond to the $k=0,1,\ldots$ terms in \ref{eq:reduced-kernel-2}.
In particular, the first order is given by 
\begin{align}
\tilde{\mathcal{K}}_{S}^{\left(1\right)}(\lambda)\bullet & =\text{Tr}_{QP}\biggl\{\mathcal{L}_{T}\tilde{\mathcal{G}_{0}}(\lambda)\mathcal{L}_{T}\bullet\otimes\hat{\rho}_{B}\biggr\},\label{eq:seq-tunneling-NZ}
\end{align}
and the second order by
\begin{align}
 & \tilde{\mathcal{K}}_{S}^{\left(2\right)}(\lambda)\bullet=\nonumber \\
 & \text{Tr}_{QP}\biggl\{\mathcal{L}_{T}\tilde{\mathcal{G}_{0}}(\lambda)\mathcal{L}_{T}\mathcal{Q}\tilde{\mathcal{G}_{0}}(\lambda)\mathcal{Q}\mathcal{L}_{T}\tilde{\mathcal{G}_{0}}(\lambda)\mathcal{L}_{T}\bullet\otimes\hat{\rho}_{B}\biggr\}.\label{eq:cotunneling-NZ}
\end{align}
The current kernel is obtained by substituting the leftmost instance
of $\mathcal{L}_{T}$ with the current operator $\hat{I}$ in these
expressions. 

 In the perturbative approach, the Josephson current appears in general
as a second-order process in $\Gamma$, as it involves the coherent
transfer of one Cooper pair (i.e. two electrons) from one superconductor
to the other through the quantum dot. As such, its full description
requires considering up to second order in the kernel. At the same
time, the formalism presented here allows one to study arbitrary interaction
strengths and gap amplitudes in an out of equilibrium situation, provided
that $\hbar\Gamma$ is the smallest energy scale in the problem. The
cases of both zero and infinite gap can be recovered in the appropriate
limits. 

Within the perturbative expansion  of \ref{eq:K-pert,eq:J-pert},
we anticipate a current of the form
\begin{equation}
I\left(t\right)=I_{0}+I_{c}\sin\left(\varphi_{0}+\Omega_{J}t\right)+I_{r}\cos\left(\varphi_{0}+\Omega_{J}t\right),\label{eq:I-de-t}
\end{equation}
with the Josephson frequency $\Omega_{J}$ defined in \ref{eq:Josephson-freq}.
The three terms in \ref{eq:I-de-t} derive from the truncation of
the harmonic series for the current in \ref{eq:current-steady-state-2}
to the terms $\bm{n}=\bm{0}$ and $\bm{n}=\left(\pm1,\mp1\right)$.
The former yields the dc term $I_{0}$. The sum of the latter has
been written as a sine term with critical current $I_{c}$, the supercurrent,
and a cosine term with amplitude $I_{r}$ which yields a phase-dependent
dissipative current~\citep{Josephson1964,Josephson1965}. Denoting
it as a ``dissipative current'' is warranted in the sense that $I_{n}\left(-V_{b}\right)=-I_{n}\left(V_{b}\right)$
for both $n=0,r$, while $I_{c}\left(-V_{b}\right)=I_{c}\left(V_{b}\right)$.
This last property ensures that there can be a non-zero current at
zero bias as long as the phase difference between the two superconductors
is not zero (modulo $\pi$). Further harmonics will appear at higher
orders in $\Gamma$, but we will not consider them in this work. 
\begin{table}
\begin{centering}
\par\end{centering}
\begin{centering}
\includegraphics{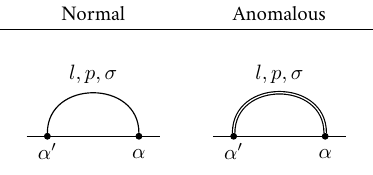}
\par\end{centering}
\caption{The two types of diagrams contributing to the first order kernel:
(left) normal and (right) anomalous. The first one accounts for quasiparticle
transport; the latter includes a Cooper pair and connects populations
to Cooper pair coherences. \label{tab:seq-tun-diags}}
\end{table}

\subsection{First order: sequential tunneling\label{subsec:Sequential-tunneling}}

As a first step, we keep only terms of order $\Gamma$ in the kernels,
yielding the so-called sequential tunneling approximation. The kernel,
given by \ref{eq:seq-tunneling-NZ}, entails two tunneling events,
each associated with an instance of the tunneling Liouvillian $\mathcal{L}_{T}$.
These events are correlated by the overall trace over the bath. In
the superconducting case, the Bogoliubov-Valatin transformation, \ref{eq:Bog-Valatin},
allows for two distinct processes. On the one hand, the trace may
be performed over two electron-like or two hole-like excitations of
opposite hermiticity (see ~\citep{PicoCortesTh2021,Siegl2023} for
details). These two contributions do not change the number of Cooper
pairs. When further summing over the Cooper pairs according to the
prescription in \ref{eq:GME-generalized-partial-tr}, these terms
combine to yield the so-called \emph{normal }kernel. It has the form

\begin{align}
\tilde{\mathcal{K}}_{N}^{\left(1\right)}\left(\bm{0};\lambda\right) & =\frac{1}{i\hbar}\sum_{l,\sigma}\sum_{p,\alpha\alpha'}\left|t_{l}\right|^{2}\alpha'\alpha\int_{-\infty}^{\infty}dE\nonumber \\
 & \times\fsop_{\sigma}^{\bar{p}\alpha'}\frac{\dos_{N,l}\left(E\right)f\left(\alpha E\right)}{E-i\hbar\mathcal{L}_{QD}+p\mu_{l}+i\hbar\lambda}\fsop_{\sigma}^{p\alpha},\label{eq:kernel-2nd-order}
\end{align}
where $f\left(E\right)=1/(1+e^{\beta E})$ is the Fermi function,
accounting for the equilibrium distribution of the quasiparticles.
We introduced the quantum dot superoperators $\fsop_{\sigma}^{p\alpha}$,
labeled by a Liouville index $\alpha$, indicating that the operator
acts from the left ($\alpha=+$, $\mathcal{\fsop}_{\sigma}^{p+}\bullet=\hat{d}_{\sigma}^{p}\bullet$)
or from the right ($\alpha=-$, $\fsop_{\sigma}^{p-}\bullet=\bullet\hat{d}_{\sigma}^{p}$).
The Fock index $p$ indicates hermiticity ($p=+$, $\hat{d}_{\sigma}^{+}=\hat{d}_{\sigma}^{\dagger}$,
$p=-$, $\hat{d}_{\sigma}^{-}=\hat{d}_{\sigma}$).  The value of $\mathcal{L}_{QD}$
in the denominator can then be determined via $i\hbar\mathcal{L}_{QD}\ket{\chi}\bra{\chi'}=(E_{\chi}-E_{\chi'})\ket{\chi}\bra{\chi'}$,
with $\hat{H}_{QD}\ket{\chi}=E_{\chi}\ket{\chi}$. 

The kernel \ref{eq:kernel-2nd-order} is similar to that of a normal
conductor with its density of states  $\dos_{0,l}$ replaced by 
\begin{equation}
\dos_{N,l}\left(E\right)=\dos_{0,l}\real\left\{ \sqrt{\frac{E^{2}}{E^{2}-|\Delta_{l}|^{2}}}\right\} ,\label{eq:DOS-func}
\end{equation}
reflecting the absence of quasiparticle excitations inside the superconducting
gap. The current kernel $\tilde{\mathcal{J}}_{N}^{\left(1\right)}\left(\bm{0};\lambda\right)$
can be obtained in a similar manner by taking $\alpha=+$ and inserting
a factor $ep\ell/2$ inside the sum in \ref{eq:kernel-2nd-order},
where $\ell=\left(-1\right)^{\delta_{lL}}$. 

On the other hand, the trace over the quasiparticles may give a nonzero
result when performed between one electron and one hole excitation
(or vice versa) of the same hermiticity, leaving one Cooper pair operator.
Summing over the Cooper pairs according to \ref{eq:kernel-elements}
yields the anomalous kernel, given by 

\begin{align}
\tilde{\mathcal{K}}_{A}^{\left(1\right)}\left(p\bm{u}_{l};\lambda\right) & =\frac{1}{i\hbar}\sum_{\sigma}\sum_{\alpha'\alpha}p\alpha'\alpha\sigma\left|t_{l}\right|^{2}\int_{-\infty}^{\infty}dE\nonumber \\
 & \times\fsop_{\sigma}^{\bar{p}\alpha'}\frac{\dos_{A,l}\left(E\right)f\left(\alpha E\right)\text{sgn}\left(E\right)}{E-i\hbar\mathcal{L}_{QD}-p\mu_{l}+i\hbar\lambda}\fsop_{\bar{\sigma}}^{\bar{p}\alpha},\label{eq:anomalous-kernel-2nd-order}
\end{align}
where $\bm{u}_{L}=\left(1,0\right),\bm{u}_{R}=\left(0,1\right)$.
Compared to the normal case, instead of the density of states as in
\ref{eq:DOS-func}, we find an anomalous density of states

\begin{equation}
\dos_{A,l}\left(E\right)=\dos_{0,l}\real\left\{ \sqrt{\frac{|\Delta_{l}|^{2}}{E^{2}-|\Delta_{l}|^{2}}}\right\} .\label{eq:DOS-func-anom}
\end{equation}
The analytic form of the integrals appearing in \ref{eq:kernel-2nd-order,eq:anomalous-kernel-2nd-order}
is provided in \ref{sec:seq-tun-integrals-1}. The kernel $\tilde{\mathcal{K}}_{A}^{\left(1\right)}\left(p\bm{u}_{l};\lambda\right)$
generates superconducting coherences $\hat{r}\left(p\bm{u}_{l}\right)$
when acting on the populations in $\hat{r}\left(\bm{0}\right)$. Due
to \ref{eq:particle-cons-coherences}, the operator $\hat{r}\left(p\bm{u}_{l}\right)$
can only have elements of the type $\ket{2}\bra{0}$ and $\ket{0}\bra{2}$
in the quantum dot sector. Note that allowing for $\Delta_{l},t_{l}\notin\mathbb{R}$
results in an additional phase factor $\exp-ip(\arg\Delta_{l}+2\arg t_{l})$
in \ref{eq:anomalous-kernel-2nd-order}. 

The processes contributing to \ref{eq:kernel-2nd-order,eq:anomalous-kernel-2nd-order}
admit a diagrammatic representation~\citep{Konig1996,GrifoniBook},
as shown in \ref{tab:seq-tun-diags}. The propagation from the initial
state (to the right) towards the final state (to the left) is represented
as a line. On this line, each fermion operator $\fsop_{\sigma}^{p,\alpha}$
is represented by a vertex, with an associated Liouville index $\alpha$.
Each operator is linked via a contraction to another one, which is
represented in the diagram via a \emph{quasiparticle arc }connecting
the two vertices. This contraction has associated indices $l,p,\sigma$
and an energy $E$. For the first order in $\Gamma$, there are only
two types of diagrams, normal and anomalous, represented in \ref{tab:seq-tun-diags}.
In the same spirit, diagrammatic representations have been provided
in \citep{Governale2008,Hiltscher2012} employing a Hilbert space
formulation. The details of the Liouville diagrammatic rules to arbitrary
order are given in \ref{sec:Diagrammatic-rules}. 

\subsection{Second order: cotunneling and pair-tunneling\label{subsec:Cotunneling}}

\begin{table}[t]
\begin{centering}
\par\end{centering}
\begin{centering}
\includegraphics{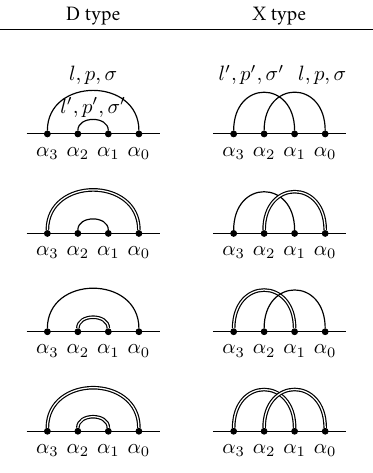}
\par\end{centering}
\caption{All classes of irreducible second order diagrams. Single arcs represent
normal contractions while double lines represent anomalous ones. The
left column displays the D type diagrams, and the right, the X type
diagrams. \label{tab:cotunneling-diags}}
\end{table}

The second order kernel is given in \ref{eq:cotunneling-NZ}. It involves
four tunneling events (i.e. one for each instance of $\mathcal{L}_{T}$).
Each kernel element is thus proportional to a bath correlator with
four quasiparticle operators. Using Wick's theorem, this correlator
can itself be written in terms of two contractions of two quasiparticle
operators. There are three non-vanishing ways to contract four operators.
The two that need to be considered are represented in the columns
of \ref{tab:cotunneling-diags}. We will refer to them as D and X
diagrams, respectively. The third type of contraction, in which the
leftmost and rightmost pairs of vertices are contracted separately,
provides a \emph{reducible diagram}, which is canceled by the $\mathcal{Q}$
projectors appearing in \ref{eq:cotunneling-NZ}. In general, any
diagram which can be divided into two by a vertical line without separating
two vertices connected by an arc is called reducible and is canceled
at this and every further perturbative level. In contrast, the D and
X diagrams are \emph{irreducible}. The differences between the D and
X diagrams in a physical sense are most apparent when considering
diagrammatic resummation techniques~\citep{Magazzu2022,GrifoniBook}. 

In the same way as we did for the first order, we can divide the correlators
according to whether the contractions are normal or anomalous. The
four possible combinations correspond to the rows of \ref{tab:cotunneling-diags}.
In order to differentiate these terms, we write $\tilde{\mathcal{K}}_{\ai'\ai}^{\left(2\right)}\left(\bm{m},\lambda\right)$
when necessary, where $\ai=N(A)$ if the arc associated with the rightmost
vertex is normal (anomalous) and equally for $\ai'$ for the other
arc. 

Kernel elements which involve two normal contractions, $\tilde{\mathcal{K}}_{NN}^{\left(2\right)}$,
do not change the Cooper pair number and thus contribute to $\tilde{\mathcal{K}}^{\left(2\right)}\left(\bm{0},\lambda\right)$
as in the first order. Furthermore, we will have kernel elements with
one normal and one anomalous arc, $\tilde{\mathcal{K}}_{AN}^{\left(2\right)},\tilde{\mathcal{K}}_{NA}^{\left(2\right)}$
corresponding to the diagrams represented in the second and third
rows of \ref{tab:cotunneling-diags}. These contribute to $\tilde{\mathcal{K}}^{\left(2\right)}\left(p\bm{u}_{l},\lambda\right)$
and thus act as a higher-order corrections to the anomalous terms
already appearing in the sequential tunneling approximation; overall
they add very little to the current almost everywhere, as can be
seen from the discussion in \ref{sec:Relevance-of-the}. 

Finally, we have the kernel elements with two anomalous arcs, $\tilde{\mathcal{K}}_{AA}^{\left(2\right)}$
, corresponding to the diagrams represented in the last row of \ref{tab:cotunneling-diags}.
Since the quantum dot can host at most two charges, one of the arcs
must create a Cooper pair, while the other must destroy one {[}see
\ref{eq:particle-cons-2}{]}. These can be further separated: if the
two anomalous arcs correspond to the same superconductor, the kernel
contributes to $\tilde{\mathcal{K}}^{\left(2\right)}\left(\bm{0},\lambda\right)$;
if each arc belongs to one distinct lead, the process contributes
to one of

\begin{equation}
\tilde{\mathcal{K}}^{\left(2\right)}[\left(-1,1\right),\lambda],\,\tilde{\mathcal{K}}^{\left(2\right)}[\left(1,-1\right),\lambda],
\end{equation}
which produce the coherent transfer of one Cooper pair from one lead
to the other. 

\section{Dc current\label{sec:dc-current}}

\begin{figure*}[t]
\begin{centering}
\includegraphics[scale=0.75]{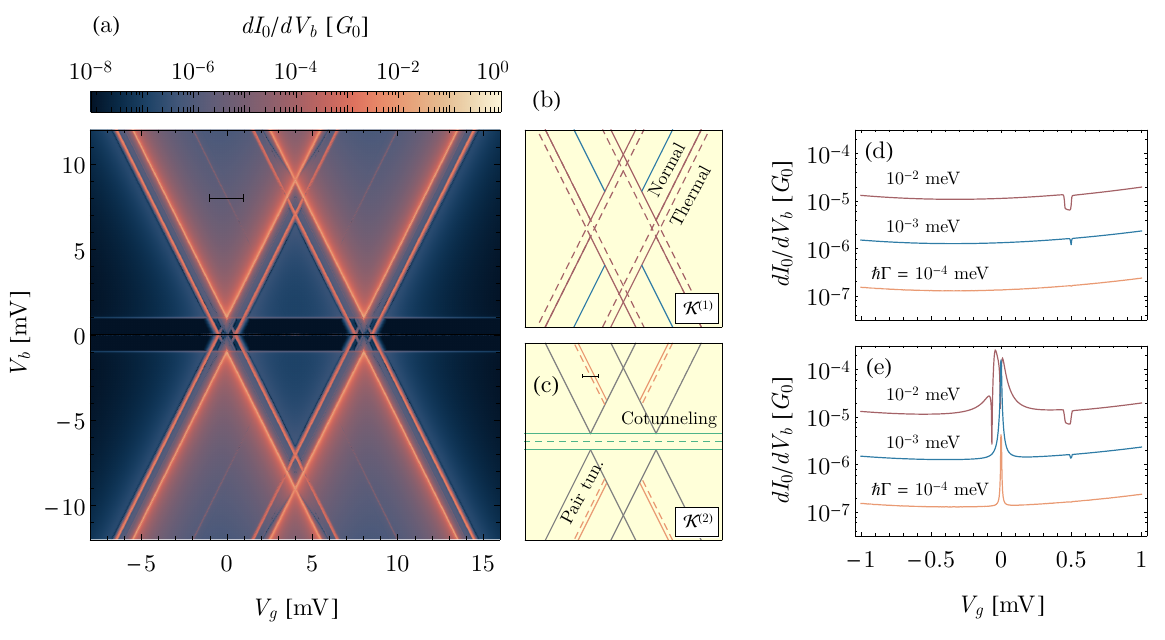}
\par\end{centering}
\raggedright{}\caption{Differential conductance $dI_{0}/dV_{b}$ up to the second perturbative
order. (a) Stability diagram for $dI_{0}/dV_{b}$ as a function of
the gate and bias voltages, $V_{b}$ and $V_{g}$, respectively. Parameters
are $\Delta=0.5\,\protect\meV,$ $U=8\,\protect\meV$, $\hbar\Gamma=10^{-3}\,\protect\meV$,
$\zeta=10^{-3}\,\protect\meV$, $k_{B}T=0.05\,\protect\meV$. A current
suppression due to the quasiparticle gap is clearly observed around
zero bias. (b,c) Sketch showing which features arise in first and
second order, respectively. In (b) normal features from quasiparticles
below the gap are represented by solid red lines, while those from
thermally activated quasiparticles above the gap are represented by
dashed lines. The resonance due to anomalous pair tunneling is represented
in blue. In (c) we have indicated elastic cotunneling (green) and
normal pair tunneling (orange) features. (d,e) Differential conductance
for $eV_{b}=U=8\,\protect\meV$ as a function of the gate voltage
without (d) and with (e) anomalous contributions and for different
values of $\hbar\Gamma$ ($10^{-2}\,\protect\meV$, $10^{-3}\,\protect\meV$,
and $10^{-4}\,\protect\meV$). The range corresponding to this figure
is indicated by a black segment in (a) and (c). \label{fig:ncurrent}}
\end{figure*}

We discuss now the features appearing in the dc current $I_{0}$ when
out of equilibrium. Specifically, we consider $eV_{b}\gg\hbar\Gamma$.
The opposite case $eV_{b}\ll\hbar\Gamma$ requires retaining all the
Cooper pair coherences in \ref{eq:GME-generalized-partial-tr}, due
to the overlap of Cooper pair resonances, as discussed in \ref{sec:Relevance-of-the}.
Moreover, we consider for numerical reasons a non-zero Dynes parameter
$\zeta$ in both the normal and the anomalous density of states, corresponding
to taking $\dos_{l,\ai}\left(E\right)\to\dos_{l,\ai}\left(E+i\zeta\right)$. 

The numerical results for the differential conductance $dI_{0}/dV_{b}$
as a function of the gate $V_{g}$ and bias $V_{b}$ voltages are
shown in \ref{fig:ncurrent}~(a) for the representative parameters
$U=8\,\meV,\Delta=0.5\,\meV,\hbar\Gamma=10^{-3}\,\meV$. In order
to understand the features seen in this figure, we distinguish between
normal and anomalous (Andreev) contributions to the dc current. In
particular, the dc current $I_{0}$ can be written as 
\begin{equation}
I_{0}=I_{QP}+I_{\text{And}},
\end{equation}
where
\begin{equation}
I_{QP}=\text{Tr}_{QD}\biggl\{\tilde{\mathcal{J}}\left(\bm{0};0\right)\hat{r}\left(\bm{0}\right)\biggr\}
\end{equation}
is the ``quasiparticle'' or ``normal'' current, and
\begin{align}
I_{\text{And}} & =\text{Tr}_{QD}\biggl\{\sum_{\bm{m}\neq\bm{0}}\tilde{\mathcal{J}}\left(-\bm{m};i\omega_{\bm{m}}\right)\hat{r}\left(\bm{m}\right)\biggr\}
\end{align}
is the ``Andreev'' or ``anomalous'' current. 

\subsection{Normal current\label{subsec:Normal-current}}

Within the perturbative expansion considered here, the normal current
can be written as
\begin{equation}
I_{QP}=I_{0,N}+I_{0,NN},
\end{equation}
where 
\begin{align}
I_{0,N} & =\text{Tr}_{QD}\{\tilde{\mathcal{J}}_{N}^{\left(1\right)}\left(\bm{0};0\right)\hat{r}\left(\bm{0}\right)\},\\
I_{0,NN} & =\text{Tr}_{QD}\{\tilde{\mathcal{J}}_{NN}^{\left(2\right)}\left(\bm{0};0\right)\hat{r}\left(\bm{0}\right)\}.
\end{align}
For these kernels, the only difference between a superconductor and
a normal metal is the non-flat density of states. Apart from the changes
to $\hat{r}\left(\bm{0}\right)$ due to the anomalous terms, this
is equivalent to the so-called ``semiconductor model'' of the superconductor~\citep{TinkhamBook}.

The normal terms produce a stability diagram which resembles the case
of a SIAM coupled to non-superconducting electrodes, once the DOS
is accounted for. In a similar way to that case, one observes areas
of reduced current due to the Coulomb interaction, the well-known
Coulomb diamonds, and areas of current flow, which take the form of
triangles, pointing at the charge degeneracy points $eV_{g}=0,U$.
In addition, in the superconducting case, the gap opens a region of
reduced current for 
\begin{equation}
-2\Delta<eV_{b}<2\Delta.\label{eq:cotunneling}
\end{equation}
On top of that, additional resonant lines running parallel to the
Coulomb diamonds and intersecting at $V_{b}=0$ can be observed. At
these resonances, the current is carried by thermally excited quasiparticles
above the gap, originating from the exponential tail of the Fermi
function~\citep{Pfaller2013}. In \ref{fig:ncurrent}~(b) we have
represented schematically the main features appearing in the first
order of perturbation theory, distinguishing normal contributions
arising from quasiparticles below the gap (solid lines) and from the
thermally excited quasiparticles (dashed lines). Moreover, we point
out the presence of a faint feature exactly at the middle point between
the normal and thermal lines, which is visible especially for voltages
satisfying \ref{eq:cotunneling}. This feature is the result of a
non-zero subgap density of states due to the Dynes parameter. Then,
$f\left(E\right)\dos_{l}\left(E\right)$ exhibits a small step at
$E=0$ due to the Fermi function and, as a result, one observes a
faint conductance peak. 

The main processes appearing at the second order are due to long-range
virtual tunneling of quasiparticles across the dot, in the form of
cotunneling, and pair tunneling. These features are represented schematically
in \ref{fig:ncurrent}~(c). 

Cotunneling leads to the presence of a significant current in the
Coulomb diamonds, larger than the contribution from the first order
term. In the case of superconducting leads, cotunneling is suppressed
for the range in \ref{eq:cotunneling} due to the unavailability of
quasiparticle excitations inside the superconducting gap. As a result,
the onset of cotunneling can be appreciated as a horizontal feature~\citep{Grove-Rasmussen2009,Ratz2014}
(i.e. independent of the gate voltage $V_{g}$) in the differential
conductance at $\pm2\Delta/e$, which can be seen clearly in \ref{fig:ncurrent}~(a).
It corresponds to the solid green lines in the scheme of \ref{fig:ncurrent}~(c).
Thermal effects also result in a finite current, albeit strongly reduced,
extending to $eV_{b}=0$ and producing a zero-bias conductance peak~\citep{Ratz2014}
{[}represented with a dashed red line in \ref{fig:ncurrent}~(c){]}. 

Quasiparticle pair tunneling involves the simultaneous transfer of
two charges between the leads. Normal pair tunneling occurs prominently
under the condition
\begin{equation}
-2eV_{g}+U+\left(-1\right)^{\delta_{l,L}}eV_{b}=2|\Delta_{l}|,\label{eq:normal-pair-tun-resonance}
\end{equation}
which can be obtained by inspection of the integrals appearing in
the pair tunneling kernel~\citep{Ratz2014}. These resonances lines
are displaced by a factor $|\Delta_{l}|$ from the pair tunneling
condition for non-superconducting leads, which corresponds to 

\begin{equation}
-2eV_{g}+U+\left(-1\right)^{\delta_{l,L}}eV_{b}=0.\label{eq:pair-tunneling-resonance}
\end{equation}
However, thermal effects extend the area of effective pair tunneling
from \ref{eq:normal-pair-tun-resonance} to \ref{eq:pair-tunneling-resonance}.
This is indicated in the scheme of \ref{fig:ncurrent}~(c) by orange
lines (again, solid for the non-thermal and dashed for the thermal
contributions).

The dc current calculated without anomalous terms is represented for
a small window around the pair tunneling resonance in \ref{fig:ncurrent}~(d),
for $eV_{b}=8\,\meV$ and different values of $\hbar\Gamma$. This
window is marked in \ref{fig:ncurrent}~(a) by a black segment. We
observe a conductance dip corresponding to normal pair tunneling at
$eV_{g}=0.5\,\meV=\Delta$ which becomes appreciable for $\hbar\Gamma>10^{-3}\,\meV$
and becomes more marked for $\hbar\Gamma=10^{-2}\,\meV$. The thermal
pair tunneling feature would appear as a peak or dip for the condition
in \ref{eq:pair-tunneling-resonance} for larger tunneling rates and/or
temperatures, but is too faint to be seen for the parameters considered
here. 

\begin{figure*}[t]
\centering{}\includegraphics[scale=0.75]{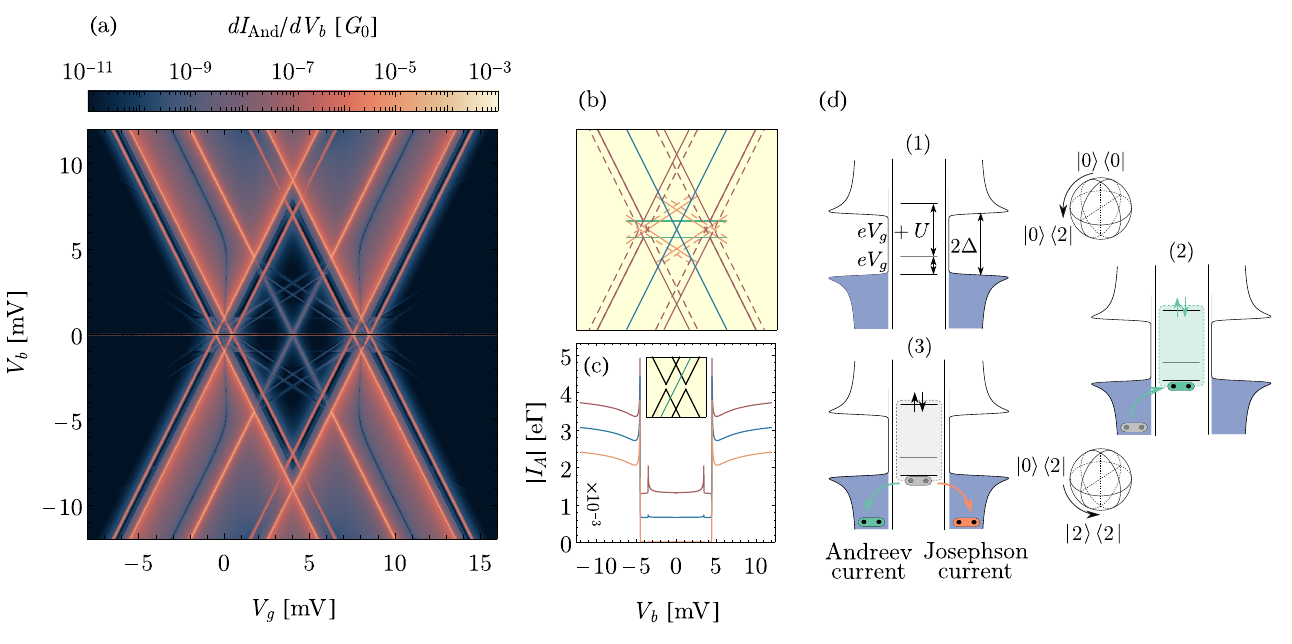}\caption{Andreev current and differential conductance up to the second perturbative
order. (a) Stability diagram depicting $dI_{\text{And}}/dV_{b}$ as
a function of the gate and bias voltages, $V_{b}$ and $V_{g}$, respectively.
(b) Scheme with the main resonances observed in panel (a). In red:
normal features (dashed lines indicate thermal processes). In blue:
pair tunneling. In orange: faint low-slope features characteristic
of anomalous transport. (c) Andreev current $I_{A}$ along the green
resonance line in the inset, for different values of $\hbar\Gamma$
according to \ref{eq:pair-tunneling-resonance} (for $l=L$). The
curves for different $\hbar\Gamma$ are displaced for visibility.
(d) Scheme of the pair transfer process. From top to bottom: (1) Energetics
of the junction. (2) The tunneling of a Cooper pair produces a Cooper
pair coherence; in the quantum dot space, this is $\ket{0}\bra{\up\dw}$.
(3) The Cooper pair is then transferred to another lead. If it comes
back to the lead it originated from, it contributes to the dc Andreev
current; on the other hand, if it tunnels to the opposite lead, it
contributes to the time-dependent Josephson effect. Parameters are
the same as in \ref{fig:ncurrent}. \label{fig:Andreev-current}}
\end{figure*}

\subsection{Andreev current\label{subsec:Andreev-current}}

We consider now the Andreev current, corresponding to the anomalous
contribution to the out of equilibrium dc current. It can be contrasted
with the Josephson current, which is time-dependent for non-zero bias.
In particular, the Andreev current occurs due to Cooper pair tunneling
between a single superconducting lead and the quantum dot, whereas
the Josephson current develops due to Cooper pair flow between two
superconducting leads. 

In the weak-coupling approximation, the Andreev current is given by
\begin{equation}
I_{\text{And}}=I_{0,A}+I_{0,AA}.
\end{equation}
The first term originates from the first-order current kernel, and
is of the form 
\begin{equation}
I_{0,A}=\text{Tr}_{QD}\{\sum_{p,l}\tilde{\mathcal{J}}_{A}^{\left(1\right)}\left(\bar{p}\bm{u}_{l};i2p\bm{u}_{l}\cdot\bm{\mu}/\hbar\right)\hat{r}\left(p\bm{u}_{l}\right)\}.\label{eq:Andreev-current-1}
\end{equation}
If the lead indices in the arguments of the kernel and the coherence
were not the same, this term would contribute to the Josephson current
{[}see also \ref{fig:Andreev-current}~(d){]}. Note that, despite
involving a first order kernel, these terms are $\propto\Gamma^{2}$
except at the pair tunneling resonances (this is discussed in more
detail in \ref{sec:Relevance-of-the}). 

The second contribution originates from the second-order anomalous
kernel, and reads
\begin{equation}
I_{0,AA}=\text{Tr}_{QD}\{\tilde{\mathcal{J}}_{AA}^{\left(2\right)}\left(\bm{0};0\right)\hat{r}\left(\bm{0}\right)\},\label{eq:Andreev-current-2}
\end{equation}
where the two anomalous arcs of $\tilde{\mathcal{J}}_{AA}^{\left(2\right)}\left(\bm{0};0\right)$
belong to the same lead. These terms are always $\propto\Gamma^{2}$. 

In \ref{fig:Andreev-current}~(a), we have represented the Andreev
conductance $dI_{\text{And}}/dV_{b}$ for the same parameters as \ref{fig:ncurrent}.
Its main features have been schematized in \ref{fig:Andreev-current}~(b).
Note that the color scale runs here from $10^{-11}$ to $10^{-3}$,
lowered by a factor $\Gamma$ compared to \ref{fig:ncurrent}~(a).
Overall, the Andreev current exhibits a similar profile as the full
dc current, with ``normal'' features along the edges of the Coulomb
diamonds, indicated by solid red lines in \ref{fig:Andreev-current}~(b).
These lines originate from the coherence peaks of the anomalous DOS,
due to the presence of a large number of energetically available quasiparticles
near the gap~\citep{Siegl2023}. Moreover, we find a peak in the
conductance along the thermal lines {[}depicted with red dashed lines
in the scheme of \ref{fig:Andreev-current}~(b){]} for identical
reasons. 

The other prominent feature is a resonant behavior along the pair
tunneling condition {[}namely \ref{eq:pair-tunneling-resonance},
marked in blue in both \ref{fig:ncurrent}~(b) and \ref{fig:Andreev-current}~(b){]}.
Under resonance, the Andreev current is $\propto\Gamma$ instead of
the usual $\propto\Gamma^{2}$, despite involving the exchange of
two particles. This scaling is shown rigorously in \ref{sec:Relevance-of-the}.
However, anomalous pair tunneling is significantly reduced inside
the $N=1$ Coulomb diamond, where the Cooper pair coherences $\hat{r}\left(\pm\bm{u}_{l}\right)$
are thermally suppressed. We have shown the Andreev current $I_{0,A}$
along one of the two resonances in \ref{fig:Andreev-current}~(c),
exhibiting both the $\propto\Gamma$ scaling and the suppression in
the first Coulomb diamond. In \ref{fig:ncurrent}~(e) we depicted
the differential conductance calculated with all of the terms in the
kernel, including the anomalous contributions for the same parameters
as in \ref{fig:ncurrent}~(d). We observe a strong feature precisely
at the pair tunneling resonance at $V_{g}=0\,\mV$, as opposed to
the resonance feature due to the normal processes, which is displaced
by $\Delta$ from the resonance condition, \ref{eq:pair-tunneling-resonance}. 

Moreover, in the current regions outside of the Coulomb diamonds,
we find a remarkable feature: a bias-independent dip in the conductance
(i.e. a vertical line)  for gate voltages corresponding to the degeneracy
points $eV_{g}=0,U$. This is the most distinct feature originating
from $I_{0,AA}$, and reflects its vanishing along these lines. At
$eV_{g}=0$, we can obtain an analytic expression for $I_{0,AA}$
in the limit $\mu_{l}\ll U$. Under this condition, the contributions
to the Andreev current from both leads cancel each other due to symmetry
and the equal populations of the empty and single occupied states.
For $eV_{g}=U$, the same logic follows due to particle-hole symmetry
of the stability diagram along the line $eV_{g}=U/2$. This is shown
in more detail in \ref{sec:Contributions-andreev}. 

Finally, we have two weaker sets of peaks, corresponding to lower
slope lines (compared to the Coulomb diamonds) and horizontal cotunneling-like
lines. These are marked respectively by orange (solid or dashed for
thermal contributions) and green lines in \ref{fig:Andreev-current}~(b).
These features are visible in \ref{fig:Andreev-current}~(a) due
to the lower range of the conductance, and correspond to effects which
scale as $\propto\Gamma^{3}$ or larger. The horizontal cotunneling-like
lines (marked in green) are due to corrections to the density operator
from the propagation kernel $\tilde{\mathcal{K}}^{\left(2\right)}$. 

The lower slope lines, indicated in orange, are reminiscent of subgap
features due to multiple Andreev reflections predicted in resonant
tunneling junctions~\citep{Johansson1999}. Here, however, due to
the inclusion of Coulomb interactions, they occur within the Coulomb
gap. While the slopes of the diamond amount to $V_{b}=\pm2V_{g}$,
the ones involving Cooper pair tunneling are given by $V_{b}=\pm2V_{g}/3$.
This reflects the energy $eV_{b}$ needed to exchange a Cooper pair
from one lead to the other. For the dc current, they appear as higher
order corrections, which are not visible in the full result of \ref{fig:ncurrent},
where cotunneling leads to an appreciable current inside the Coulomb
diamond that masks these peaks. Processes involving the transfer
of more Cooper pairs would yield even lower slopes: $\pm2/(2n+1),\,n\in\mathbb{N}$,
but are not observed in the weak coupling limit. Nonetheless, we remark
that, in the calculations presented here, only a few effects of order
$\Gamma^{3}$ or higher are accounted for. As a result, this discussion
can only be understood qualitatively. 

\begin{figure*}[t]
\begin{centering}
\includegraphics[scale=0.75]{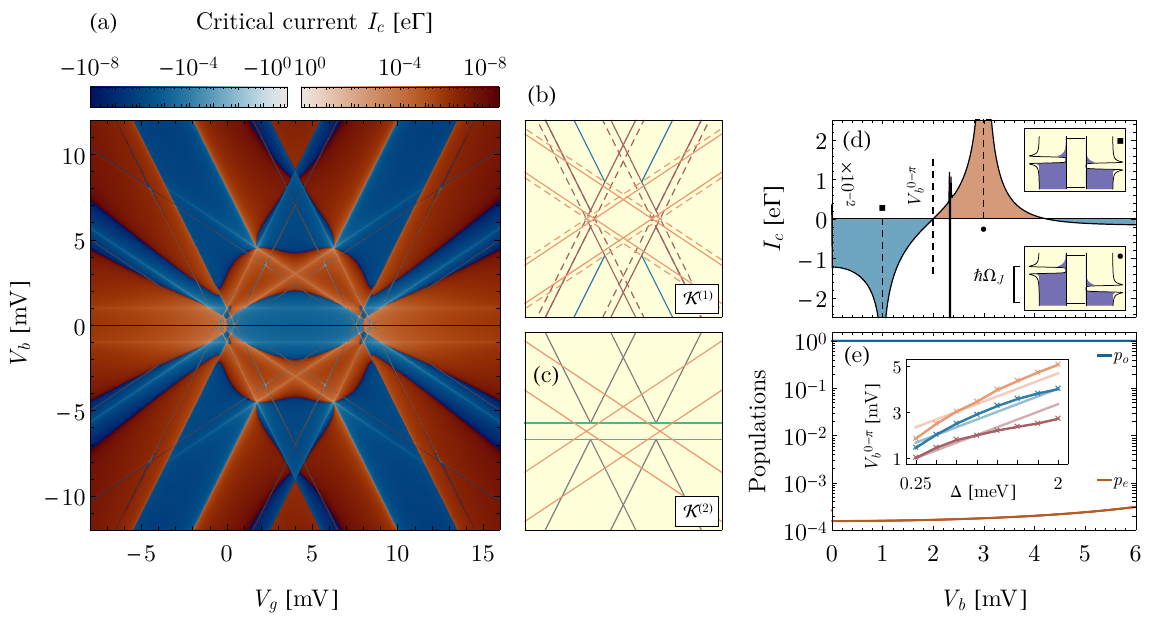}
\par\end{centering}
\caption{Critical current of the SIAM. (a) Stability diagram depicting $I_{c}$
in logarithmic scale as a function of the gate and bias voltages,
$V_{b}$ and $V_{g}$, respectively. Noticeable is the alternance
of regions with positive (red) and negative (blue) sign of the critical
current. (b) The main features originating from the first order kernel.
In red: lines reproducing the Coulomb diamond structure (dashed lines
indicate thermal features). In blue: pair tunneling resonances due
to superconducting coherences. In orange: low slope lines characteristic
of anomalous transport. (c) The unique contributions due to the second
order kernel. In green: cotunneling-like lines; i
orange: low slope lines.  (d) Critical current for $eV_{g}=U/2$.
Inset: scheme of the energetics at the cotunneling-like
(square) and low slope lines (dot). (e) Population of the odd (red)
and even (blue) sectors. Inset: the $0-\pi$ transition bias as a
function of the gap $\Delta$ for $U=4$ (red), $U=8$ (blue) and
$U=12$ (orange). The lines with lighter colors correspond to the
heuristic \ref{eq:phsym-0pitrans}. Unless noted, parameters are as
in \ref{fig:ncurrent}. \label{fig:supercurrent-full}}
\end{figure*}

We further discuss the two contributions to the Andreev current $I_{0,A}$
and $I_{0,AA}$ separately in \ref{sec:Contributions-andreev}. In
particular, inside the Coulomb diamonds, the two contributions cancel
each other as they come with similar magnitude but opposite sign.

\section{Josephson current\label{sec:Josephson-current}}

We move now to the time-dependent terms. Like we did with the Andreev
current, we can differentiate contributions coming from the first
order kernel, which have the form
\begin{equation}
\tilde{\mathcal{J}}_{A}^{\left(1\right)}\left(\mp\bm{u}_{\bar{l}};\lambda\right)\hat{r}\left(\pm\bm{u}_{l}\right),\label{eq:J-extra}
\end{equation}
with $\bar{l}$ denoting the opposite lead to $l$ (for conciseness,
here and onward we do not write explicitly the value of $\lambda$
unless necessary). We denote the current originating from these terms
as $I_{c,A}$ and $I_{r,A}$, respectively for the critical and dissipative
current. Note that, even if the current kernel is of first order,
the contribution to the current will be of order $\Gamma^{2}$ outside
of the pair-tunneling resonances. 

Moreover, we have terms coming from the second order current kernel,
of the form
\begin{equation}
\tilde{\mathcal{J}}_{AA}^{\left(2\right)}[\left(-1,1\right);\lambda]\hat{r}\left(\bm{0}\right),\tilde{\mathcal{J}}_{AA}^{\left(2\right)}[\left(1,-1\right);\lambda]\hat{r}\left(\bm{0}\right).\label{eq:kernel-anom-sc}
\end{equation}
We denote the resulting current terms by $I_{c,AA}$ and $I_{r,AA}$
for the critical and dissipative currents, respectively. The expressions
for the kernels in second order are quite complex and we describe
them in \ref{sec:Cotunneling-integrals}. Other contributions exist,
such as those coming from terms like $\tilde{\mathcal{J}}_{NN}^{\left(2\right)}\left(\bm{0};\lambda\right)\hat{r}[(-1,1)]$.
Since these are of higher order in $\Gamma$, we do not discuss them
here. 

\begin{figure*}[t]
\begin{centering}
\includegraphics[scale=0.75]{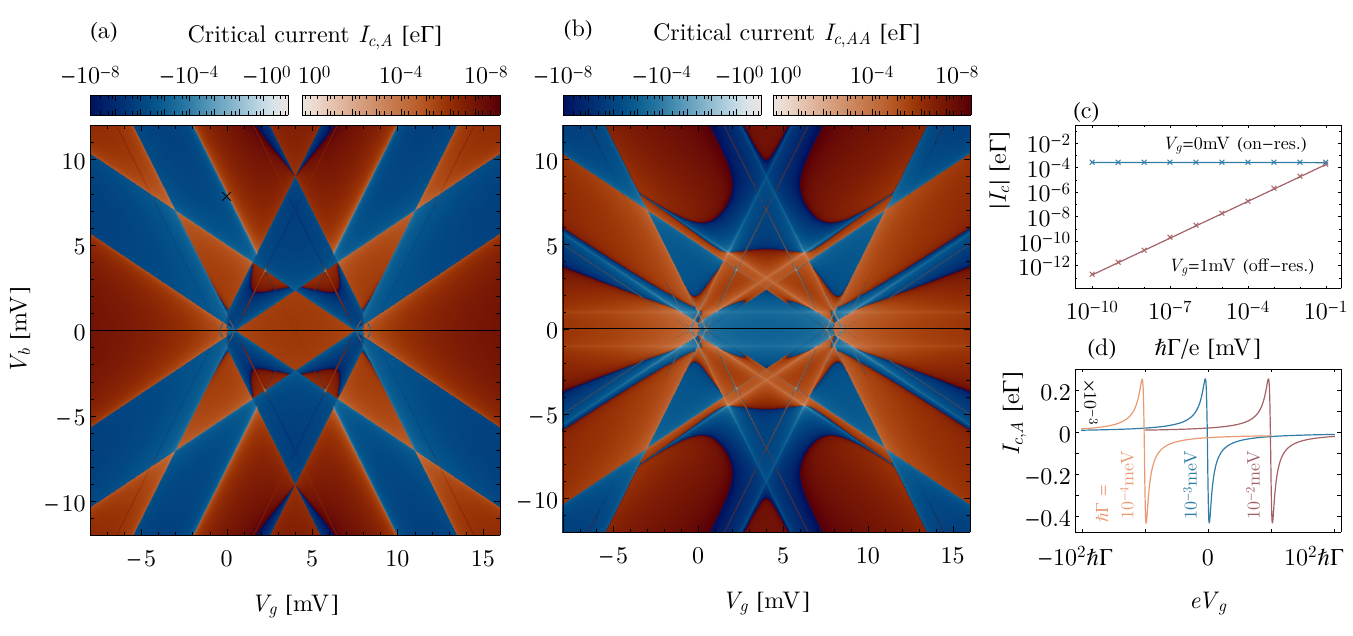}
\par\end{centering}
\caption{Contributions to the critical current originating from the first and
second order current kernels. (a) Critical current $I_{c,A}$ originating
from the first order kernel, as a function of the gate and bias voltages,
$V_{b}$ and $V_{g}$, respectively. Parameters are the same as \ref{fig:ncurrent}.
Sign is marked by color (red for positive and blue for negative sign).
(b) Same, for $I_{c,AA}$, arising from the second order kernel $I_{c,AA}$.
(c) Scaling of the critical current in a Cooper pair resonance ($V_{g}=0,V_{b}=8\,\protect\mV$,
blue) and out of resonance ($V_{g}=-1\,\protect\mV,V_{b}=8\,\protect\mV$,
red) for $\hbar\Gamma=10^{-3}\,\protect\meV$. (d) The pair tunneling
resonance at $V_{b}=8\,\protect\mV$ for $\hbar\Gamma=10^{-4}$, $10^{-3}$,
and $10^{-2}\,\protect\meV$ , displaced for clarity. The range of
$eV_{g}$ in the x axis has also been scaled by the value $\hbar\Gamma$,
showing that both the current and the width of the resonance are proportional
to $\hbar\Gamma$, \label{fig:supercurrent-contrib}}
\end{figure*}

\subsection{Critical current\label{subsec:Critical-current}}

We have depicted the critical current  as a function of the gate
and bias voltages in \ref{fig:supercurrent-full}~(a). We have represented
its magnitude logarithmically as well as its sign by the color of
the scale (red for positive and blue for negative sign). The features
in the figure coming from the first order kernel are sketched in \ref{fig:supercurrent-full}~(b),
while the ones due to the second order kernel are shown in \ref{fig:supercurrent-full}~(c).
Moreover, the two contributions $I_{c,A}$ and $I_{c,AA}$ are shown
separately in \ref{fig:supercurrent-contrib}~(a) and~(b). We discuss
now these features. 

First, the Coulomb-diamond profile can be recognized in \ref{fig:supercurrent-full}~(a).
Along these lines, the proximity effect is amplified by the coherence
peaks of the DOS, in a similar way to what occurs for the Andreev
current. Likewise, we observe peaks in the Josephson current along
the Cooper pair resonances {[}\ref{eq:pair-tunneling-resonance}{]},
marked in blue in \ref{fig:supercurrent-full}~(b). These peaks are
due to $I_{c,A}$, and thus are observed in \ref{fig:supercurrent-contrib}~(a).
The proximity effect here is the largest as the unoccupied and doubly
occupied states in the dot are resonant with the Cooper pairs of one
of the leads, as mentioned above and discussed in detail in \ref{sec:Relevance-of-the}.
There, we show that the Josephson current along these resonances is
$\propto\Gamma$ instead of the usual $\propto\Gamma^{2}$. This is
demonstrated numerically in \ref{fig:supercurrent-contrib}~(c) and~(d).
In particular, (d) shows that the width of the resonance is also proportional
to $\Gamma$. Inside the central Coulomb diamond the Cooper pair resonances
are thermally suppressed and the corresponding features are no longer
seen.

The characteristic feature of the critical current is the presence
of low slope lines, similar to the Andreev current. These are marked
in orange in \ref{fig:supercurrent-full}~(b) and ~(c). For the
Andreev current these lines correspond to higher order processes and
are barely noticeable in the full dc current of \ref{fig:ncurrent};
however, for the critical current the low slope lines are a leading
order feature. Inspecting the second order kernel $\tilde{\mathcal{J}}_{AA}^{\left(2\right)}[\left(-1,1\right);\lambda]$,
as given in \ref{sec:Cotunneling-integrals} {[}i.e. \ref{eq:anomalous D kernel}{]},
one observes a characteristic factor $3eV_{b}/2$ in the denominators,
which results in a low slope. As mentioned in \ref{subsec:Andreev-current},
this reflects the energy $eV_{b}$ needed to exchange a Cooper pair
from one lead to the other. 

Finally, we observe gate-independent horizontal lines mirroring the
cotunneling lines observed and discussed in the dc current. The origin
of these lines is similarly the long-range transfer of charge from
lead to lead with only virtual occupation of the dot. They appear
only in $I_{c,AA}$, as can be seen in \ref{fig:supercurrent-contrib}~(b). 

The critical current exhibits a complex structure when it comes to
the sign. For low bias voltages we can connect these findings to known
results for the interacting SIAM at zero bias~\citep{Glazman1989,Governale2008}.
We observe clearly a $0-\pi$ transition occurring near zero bias
upon sweeping the gate voltage. It is signaled by a change of the
sign of the supercurrent in the central Coulomb diamond compared to
the neighboring diamonds with even occupancy. This change of sign
remarkably extends to bias voltages $|eV_{b}|\lesssim3\Delta/2$.
Here, it is useful to distinguish the behavior of $I_{c,A}$ from
that of $I_{c,AA}$. For $I_{c,A}$, the critical current is mostly
positive in the central Coulomb diamond, near zero bias, except for
two small pockets near the degeneracy points $eV_{g}=0,U$. Previous
results in the low temperature limit $\beta|\Delta_{l}|\to\infty$
indicate that the contribution from the first order kernel to the
critical current is always positive~\citep{Governale2008}. However,
thermal effects allow for a change of sign to occur even at this level
of approximation due to quasiparticle-assisted processes. In contrast,
$I_{c,AA}$ is negative in the first Coulomb diamond for low bias,
except near the resonances close to the degeneracy points.

Inside the central Coulomb diamond, we observe another change in sign
in the full critical current as the voltage bias is increased. This
sign change accompanies the low slope lines at first, close to the
degeneracy points. As the bias voltage is increased further, the current
goes back to negative for $eV_{b}\gtrsim U/2$. 

We have depicted the current for $eV_{g}=U/2$ in \ref{fig:supercurrent-full}~(d),
as well as the populations of the odd and even sectors, namely
\begin{align*}
p_{o} & =\sum_{\sigma}\bra{\sigma}\hat{r}\left(\bm{0}\right)\ket{\sigma},\\
p_{e} & =\bra{0}\hat{r}\left(\bm{0}\right)\ket{0}+\bra{2}\hat{r}\left(\bm{0}\right)\ket{2},
\end{align*}
in \ref{fig:supercurrent-full}~(e). Note, in particular, that the
$0-\pi$ transition at $eV_{b}=2\,\meV$, seen in \ref{fig:supercurrent-full}~(c),
is not accompanied with a change of the parity of the ground state.
Instead, we can attribute this change to the competition between two
processes: the feature occurring at $eV_{b}=1\,\meV$ (indicated with
a square), with negative sign, and the one occurring at $eV_{b}=3\,\meV$
(marked with a dot), with positive sign. The former is the gate-independent
cotunneling-like feature at $eV_{b}=2\Delta$, while the second originates
from the low slope lines which meet in the center of the odd Coulomb
diamond at $eV_{b}=2(U/2+\Delta)/3$. Here, the $0-\pi$ transition
occurs precisely at the middle point between these two resonances.
Following this logic, we can obtain a heuristic expression for the
first $0-\pi$ transition at $eV_{g}=U/2$, which will occur for values
of the bias voltage given by 
\begin{equation}
eV_{b}^{0-\pi}\simeq U/6+4\Delta/3,\quad eV_{g}=U/2.\label{eq:phsym-0pitrans}
\end{equation}
We have calculated numerically $V_{b}^{0-\pi}$ for different values
of $U$ and $\Delta$, shown in the inset of \ref{fig:supercurrent-full}~(e).
The heuristic curves of \ref{eq:phsym-0pitrans} have been represented
with lighter colors. As can be seen there, they approximate qualitatively
the true transition points. 

Moreover, we have sketched the energetics corresponding to the two
resonances in the insets of \ref{fig:supercurrent-full}~(d), showcasing
the distinct physics corresponding to the two resonances. The gate-independent
cotunneling-like feature occurs as the peaks of the two DOS functions
align. Meanwhile, the low slope line corresponds to a Cooper pair-assisted
process, where the peak of the electron (empty) part of the source
DOS is separated by $\hbar\Omega_{J}$ from the chemical potential
of the dot at the 0-1 transition. 

In a similar way, in the $0$ and $2$ charge Coulomb diamonds, the
current around the low slope lines is of opposite sign to the rest
of the diamond. Outside of the diamonds, $I_{c,A}$ undergoes a change
of sign as the pair tunneling resonances are crossed, which is not
present in $I_{c,AA}$ (having no intrinsic pair tunneling features).
This change in sign can be understood from the fact that the contribution
to $I_{c,A}$ from lead $l$ is proportional to $\approx2eV_{g}+U-2\mu_{l}$
{[}see \ref{eq:solution-harmonic}{]}.

We note that, due to numerical errors, the resonance lines in the
critical current usually exhibit small regions of the opposite sign,
e.g. in correspondence with the normal slope thermal lines, visible
in both~\ref{fig:supercurrent-full} and~\ref{fig:supercurrent-contrib}.
We associate these localized sign issues to convergence problems when
dealing with the strongly peaked density of states. We refer to \ref{sec:Cotunneling-integrals}
for a more detailed discussion of the numerical implementation.

\subsection{Dissipative contribution \label{subsec:PDDC}}

\begin{figure}[t]
\begin{centering}
\includegraphics[scale=0.75]{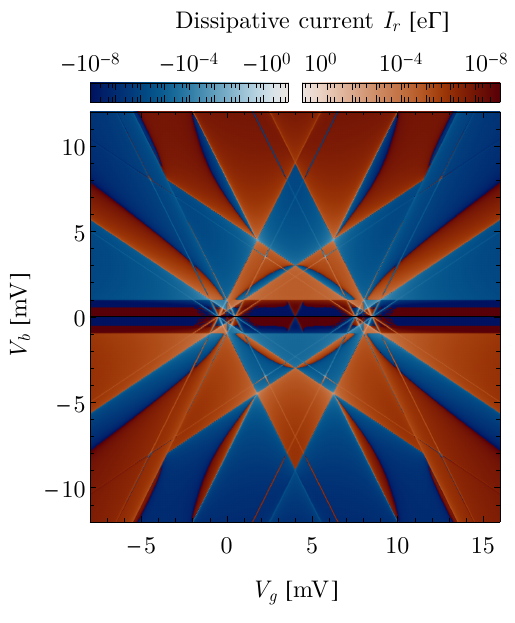}
\par\end{centering}
\caption{Dissipative contribution as a function of the gate $V_{g}$ and bias
$V_{b}$ voltages in logarithmic scale for the same parameters as
in \ref{fig:supercurrent-full}. The sign of the current is color
coded (red for positive sign and blue for negative sign). \label{fig:PDDC}}
\end{figure}

The stability diagram of the dissipative contribution $I_{r}$ is
presented in \ref{fig:PDDC}. We employ the same logarithmic scale
as for $I_{c}$, with the sign indicated by the color. The features
of the dissipative current are similar to those of $I_{c}$. Nonetheless,
we outline some differences. First, we observe horizontal cotunneling-like
features. Like for the dc current, and unlike for $I_{c}$, we observe
a stark reduction in the amplitude of the current for voltages $|eV_{b}|<2\Delta$.
As for the critical and Andreev currents, we observe low slope lines,
doubled as well into normal and thermal lines. 

Regarding the sign of $I_{r}$, we notice a similar change of sign
in the central Coulomb diamond for $|eV_{b}|>2\Delta$. Crucially,
the dissipative current  is odd in the bias voltage, in a similar
way to the dc current. Hence, it vanishes at zero bias, leaving only
the sine term in \ref{eq:I-de-t}. We observe sign changes occurring
together with the presence of low slope lines. Here, unlike for $I_{c}$,
the sign change occurs along \textemdash{} instead of around \textemdash{}
the resonance. Since $I_{r}$ is strongly suppressed for $|eV_{b}|<2\Delta$,
we do not expect the sign changes in this region to be remarkable.

\subsection{Josephson radiation\label{subsec:Josephson-radiation}}

\begin{figure}
\begin{centering}
\includegraphics[scale=0.75]{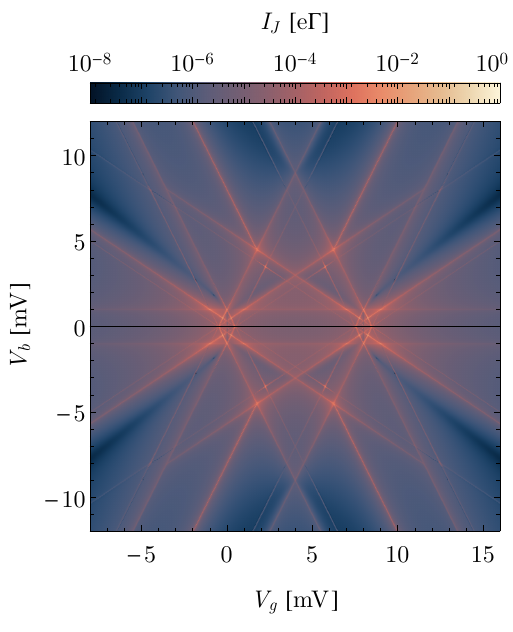}
\par\end{centering}
\caption{Josephson radiation amplitude $I_{J}$ as a function of the gate and
bias voltages in logarithmic scale for the same parameters as in \ref{fig:supercurrent-full}.
\label{fig:Josephson-radiation-amplitude}}
\end{figure}

The time-dependent components of the current can be hard to detect
in experimental circumstances. A viable way to do this is the measurement
of the Josephson radiation~\citep{Deacon2017,Watfa2021}. Hence,
we close this section by considering the features that can be accessed
via radiative measurements. In particular, we may consider an experimental
setup similar to the one in \citep{Watfa2021}. There, a superconductor-insulator-superconductor
(SIS) junction was employed as the detector, coupled to the S-QD-S
junction via a waveguide resonator. The photo-assisted tunneling current
through the SIS could be related to the amplitude of the Josephson
current as $I_{\text{PAT}}=C(V_{b})I_{J}^{2}$, where 
\begin{equation}
I_{J}=\sqrt{I_{c}^{2}+I_{r}^{2}},
\end{equation}
and $C(V_{b})$ is a voltage-dependent factor. It accounts for the
current-voltage characteristic of the SIS junction and the impedance
of the resonant circuit. We have shown $I_{J}$ as a function of the
gate and voltage bias in \ref{fig:Josephson-radiation-amplitude}.
Most of the features present in the Josephson radiation have already
been discussed in \ref{subsec:Critical-current,subsec:PDDC}. In particular,
one sees both cotunneling-like features and low slope lines. The former
are faint but observable in the results  of \citep{Watfa2021}.

\section{Conclusion\label{sec:Discussion}}

In this work we have presented a full transport theory for quantum
dot-based Josephson junctions. A particle-conserving theory of superconductivity
in the particle number space was employed, which allows one to account
for particle transfer in situations out of equilibrium. In this formulation,
the Josephson effect arises naturally from the coherent dynamics of
the Cooper pairs. 

The theory is tailored to weakly coupled systems, whereby all processes
up to second order in the hybridization strength were included. This
encompasses sequential and virtual transfer of both quasiparticles
and Cooper pairs. Specifically, we distinguished between the normal
and the anomalous contributions to tunneling. The former  conform
to the usual ``semiconductor model'' of superconductivity~\citep{TinkhamBook}.
The latter accounts for coherent Cooper pair transport and yields
both an anomalous contribution to the dc current, the so-called Andreev
current, as well as the Josephson current. 

For the dc current, we have identified characteristic features at
the Cooper pair-induced resonances, in which the quantum dot energetically
favors the appearance of coherences involving both Cooper pairs and
dot electrons. At these resonances, the current scales linearly with
the tunneling rates, instead of the usual dependence on the square,
and despite reflecting the transfer of two charges. 

For the Josephson current, we obtained the full dependence of the
critical current on the gate and on the bias voltage between the two
superconducting leads. By studying its sign we have identified the
extension of the well-known $0-\pi$ transition to out of equilibrium
situations. We found, in particular, a rich behavior of the current
(and its sign). While for the case of zero bias the change in sign
can be related directly to the parity of the ground state, for non-zero
bias these sign changes reflect a complex interplay of processes.
For example, inside the Coulomb diamonds they arise from the competition
between gate-independent cotunneling-like transport and Cooper-pair
assisted tunneling, appearing as horizontal and low slope lines in
the stability diagram, respectively. We further present results for
the dissipative current and the Josephson radiation, which exhibit
similar features to the critical current. 

The dc current has been measured in many different setups in the past,
most notably in nanowire and carbon nanotube junctions~\citep{DeFranceschi2010}.
Furthermore, measuring the Josephson radiation~\citep{Deacon2017,Watfa2021}
allows access to information about the time-dependent quantities.
From the theoretical point of view, the formalism is extremely adaptative.
While we considered the particularities of a single quantum dot junction,
the model can be extended to arbitrary interacting nanojunctions directly.

\begin{acknowledgments}
We thank Julian Siegl for discussion and help on this work, and Christoph
Br�ckner for help with the numerical simulations. J.P.-C. and M.G.
acknowledge DFG funding through Project B04 and A.D. through Project
B02 of SFB 1277 Emerging Relativistic Phenomena in Condensed Matter.
G. P. is supported by Spain\textquoteright s MINECO through Grant
No. PID2020-117787GB-I00 and by CSIC Research Platform PTI-001. 
\end{acknowledgments}

\appendix

\section{Stationary solutions of the GME\label{sec:NZE-formalism}}

We sketch the derivation of the GME, \ref{eq:Nakajima-Zwanzig-eq},
within the Nakajima-Zwanzig formalism~\citep{Nakajima1958,Zwanzig1960},
focusing on the time-dependent solutions. 

The full density operator, including quantum dot and leads, is a solution
of the Liouville-von Neumann equation

\begin{equation}
\partial_{t}\hat{\rho}\left(t\right)=-\frac{i}{\hbar}\bigl[\hat{H},\hat{\rho}\left(t\right)\bigr]=\mathcal{L}\hat{\rho}\left(t\right),\label{eq:Liouville-von-Neumann-eq}
\end{equation}
The Nakajima-Zwanzig projector is now introduced
\begin{equation}
\mathcal{P}\bullet=\text{Tr}_{B}\{\bullet\}\otimes\hat{\rho}_{B}=\left(1-\mathcal{Q}\right)\bullet,
\end{equation}
with an, in principle, arbitrary reference density operator $\hat{\rho}_{B}$.
Provided that $\mathcal{L}_{B}\hat{\rho}_{B}=0$, we can write \ref{eq:Liouville-von-Neumann-eq}
as an equation only for $\mathcal{P}\hat{\rho}\left(t\right)$ as

\begin{align}
\partial_{t}\mathcal{P}\hat{\rho}\left(t\right) & =\mathcal{P}\mathcal{L}_{T}\tilde{\mathcal{G}}_{Q}\left(t\right)\mathcal{Q}\hat{\rho}\left(0\right)\label{eq:Nakajima-Zwanzig}\\
 & +\mathcal{L}_{S}\mathcal{P}\hat{\rho}\left(t\right)+\int_{0}^{t}ds\mathcal{K}_{\mathcal{P}}\left(t-s\right)\mathcal{P}\hat{\rho}\left(s\right),\nonumber 
\end{align}
{[}compare \ref{eq:Nakajima-Zwanzig-eq}{]} where we have defined
\begin{align}
\mathcal{K}_{\mathcal{P}}\left(t-s\right) & =\mathcal{P}\mathcal{L}_{T}\bar{\mathcal{G}}_{\mathcal{Q}}\left(t-s\right)\mathcal{L}_{T}\mathcal{P},\label{eq:kernel}\\
\bar{\mathcal{G}}_{\mathcal{Q}}\left(t-s\right) & =\exp\left(\mathcal{L}_{S}+\mathcal{L}_{B}+\mathcal{Q}\mathcal{L}_{T}\mathcal{Q}\right)\left(t-s\right).
\end{align}
We moreover assume that $\mathcal{Q}\hat{\rho}\left(0\right)=0$,
allowing us to neglect the first term in the right-hand side of \ref{eq:Nakajima-Zwanzig}.

For the time-independent case considered here, we proceed by Laplace
transforming \ref{eq:Nakajima-Zwanzig}, yielding
\begin{align}
(\lambda-\mathcal{L}_{S})\lapt\left\{ \mathcal{P}\hat{\rho}\right\} \left(\lambda\right) & -\mathcal{P}\hat{\rho}\left(0^{+}\right)\label{eq:Laplace-eq}\\
= & \lapt\left\{ \mathcal{K}_{\mathcal{P}}\right\} \left(\lambda\right)\lapt\left\{ \mathcal{P}\hat{\rho}\right\} \left(\lambda\right),\nonumber 
\end{align}
with $\lapt\left\{ f\right\} \left(\lambda\right)=\int_{0}^{\infty}dte^{-\lambda t}f\left(t\right)$
denoting Laplace transformation. The time-domain solution can be recovered
through Mellin's inverse formula
\begin{equation}
\mathcal{P}\hat{\rho}\left(t\right)=\frac{1}{2\pi i}\int_{\gamma-i\infty}^{\gamma+i\infty}d\lambda\,e^{\lambda t}\lapt\left\{ \mathcal{P}\hat{\rho}\right\} \left(\lambda\right),\label{eq:Mellin-inverse-formula}
\end{equation}
where $\gamma$ is a real number such that all poles of $\lapt\left\{ \mathcal{P}\hat{\rho}\right\} \left(\lambda\right)$
lie to the left of the $\real\{\lambda\}=\gamma$ line in the complex
plane. This integral can be calculated with the use of the residue
theorem, employing a semicircular contour in the left complex half-plane.
Then, provided that there are only first order poles in the contour,
we recover directly the time evolution

\begin{equation}
\mathcal{P}\hat{\rho}\left(t\right)=\sum_{\lambda_{i}}e^{\lambda_{i}t}\hat{r}_{S}\left(\lambda_{i}\right)\otimes\hat{\rho}_{B},\label{eq:time-dep-Prho-after-Mellin}
\end{equation}
where $\hat{r}_{S}\left(\lambda_{i}\right)$, defined in \ref{eq:residues-lambda},
are the residues at the poles $\lambda_{i}$. Upon taking the trace
over the bath, we can define the kernel
\begin{equation}
\tilde{\mathcal{K}}_{S}(\lambda)\bullet=\text{Tr}_{B}\{\mathcal{L}_{T}\lapt\{\bar{\mathcal{G}}_{\mathcal{Q}}\}(\lambda)\mathcal{L}_{T}\bullet\otimes\hat{\rho}_{B}\},\label{eq:reduced-kernel}
\end{equation}
which has the expanded form of \ref{eq:reduced-kernel-2}. Then,
\ref{eq:Laplace-eq} can be written as 
\begin{align}
(\lambda-\mathcal{L}_{S})\hat{r}_{S}\left(\lambda_{i}\right)-\text{Tr}_{B}\{\mathcal{P}\hat{\rho}\left(0^{+}\right)\} & =\tilde{\mathcal{K}}_{S}\left(\lambda\right)\hat{r}_{S}\left(\lambda_{i}\right),\label{eq:Laplace-eq-1}
\end{align}
The poles $\lambda_{i}$ can be calculated by taking the limit $\lambda\to\lambda_{i}$
in \ref{eq:Laplace-eq-1}, yielding
\begin{align}
[\mathcal{L}_{S}-\lambda_{i}+\tilde{\mathcal{K}}_{S}(\lambda_{i})]\hat{r}_{S}\left(\lambda_{i}\right) & =0.\label{eq:constituyent-equation-2}
\end{align}
The $\lambda_{i}$ are the solutions to the non-linear eigenvalue
equation
\begin{equation}
\text{det}\{\mathcal{L}_{S}-\lambda+\tilde{\mathcal{K}}_{S}(\lambda)\}=0,\label{eq:constituyent-equation}
\end{equation}
Here the determinant is understood to be of the linear application
$\mathcal{L}_{S}-\lambda+\tilde{\mathcal{K}}_{S}(\lambda)$ acting
on the space of operators. 

Likewise, the Laplace transform of the current
\begin{equation}
I\left(\lambda\right)=\text{Tr}\left\{ \lapt\{\mathcal{J}\}\left(\lambda\right)\lapt\left\{ \mathcal{P}\hat{\rho}\right\} \left(\lambda\right)\right\} ,\label{eq:current-Laplace-transformed}
\end{equation}
can be transformed back to real time using Mellin's inverse formula,
yielding \ref{eq:current-Laplace-transformed-1}, with the current
kernel defined in analogy to \ref{eq:reduced-kernel}, namely
\begin{equation}
\tilde{\mathcal{J}}_{S}(\lambda)\bullet=\text{Tr}_{B}\{\hat{I}\lapt\{\bar{\mathcal{G}}_{\mathcal{Q}}\}(\lambda)\mathcal{L}_{T}\bullet\otimes\hat{\rho}_{B}\},\label{eq:current-kernel}
\end{equation}
where $\hat{I}$ is defined above \ref{eq:current-red}. 

\section{GME for the quantum dot and Cooper pairs\label{sec:NZE-sc}}

We particularize the Nakajima-Zwanzig formalism to the case of a central
system comprising both the quantum dot states and the Cooper pairs
of the leads. The action of the current kernel can be written in the
following manner, separating explicitly the action on the QD and CP
sectors \begin{widetext}

\begin{align}
\tilde{\mathcal{J}}_{S}\left(\lambda\right)\hat{r}_{S}\left(\lambda\right)= & \sum_{\delta\bm{M}}\sum_{\Delta\bm{M}}\sum_{\bm{m}}\sum_{\bm{M}}\left(\hat{S}_{L}^{\dagger}\right)^{\delta M_{L}}\left(\hat{S}_{R}^{\dagger}\right)^{\delta M_{R}}\ket{\bm{M}+\bm{m}}\bra{\bm{M}}\left(\hat{S}_{R}^{\dagger}\right)^{\Delta M_{R}}\left(\hat{S}_{L}^{\dagger}\right)^{\Delta M_{L}}\nonumber \\
\times & \tilde{\mathcal{J}}\left(\delta\bm{M},\Delta\bm{M};\lambda+i\omega_{\bm{m}}\right)\hat{r}\left(\bm{m},\bm{M};\lambda\right),\label{eq:action-of-Jl}
\end{align}
where $(\hat{S}_{l}^{\dagger})^{-m}$ is $(\hat{S}_{l})^{m}$, $\omega_{\bm{m}}$
is defined in \ref{eq:omega_n}, and we introduce two difference vectors
$\delta\bm{M}$ and $\Delta\bm{M}$ . \end{widetext}

Here, $\tilde{\mathcal{J}}\left(\delta\bm{M},\Delta\bm{M};\lambda\right)$
is a superoperator in the quantum dot space only. In order to arrive
to an expression for it, we have to evaluate the effect of the kernel
$\tilde{\mathcal{J}}_{S}\left(\lambda\right)$ on the Cooper pair
sector. This action comes in two forms. First, through the presence
of the Cooper pair operators, and second through the action of $\mathcal{L}_{CP}$,
present in the free propagators {[}see \ref{eq:propagator}{]}, which
needs to be resolved. In order to do so, we make use of $i\hbar\mathcal{L}_{CP}\hat{S}_{l}^{p}=2p\mu_{l}\hat{S}_{l}^{p}$
, to move the Cooper pair operators to the left, as in \ref{eq:action-of-Jl}.
By this, we have included a shift $i2\bm{m}\cdot\bm{\mu}/\hbar$ in
the argument of the kernel in Laplace space. This accounts for the
action of $\mathcal{L}_{CP}$ when applied to a term $\ket{\bm{M}+\bm{m}}\bra{\bm{M}}$
such as the one appearing in \ref{eq:action-of-Jl}. We could incorporate
this as a shift in the Laplace variable $\lambda$ because $\lambda$
and $\mathcal{L}_{CP}$ always appear together in the propagators. 

The current can be written in terms of the $\tilde{\mathcal{J}}\left(\delta\bm{M},\Delta\bm{M};\lambda\right)$
of \ref{eq:action-of-Jl} as\begin{widetext}
\begin{align}
I^{\left(\lambda\right)}=\text{Tr}_{QD} & \biggl\{\sum_{\Delta\bm{M}}\sum_{\bm{m}}\sum_{\bm{M}}\tilde{\mathcal{J}}\left(-\bm{m}-\Delta\bm{M},\Delta\bm{M};\lambda+i\omega_{\bm{m}}\right)\hat{r}\left(\bm{m},\bm{M};\lambda\right)\biggr\}.\label{eq:current-developed-1-1}
\end{align}
\end{widetext}Since the kernel term is independent of the index $\bm{M}$,
we can move the summation over $\bm{M}$ to the density operator and
introduce the simplified notation
\begin{align}
I^{\left(\lambda\right)}=\text{Tr}_{QD} & \biggl\{\sum_{\bm{m}}\tilde{\mathcal{J}}\left(-\bm{m};\lambda+i\omega_{\bm{m}}\right)\hat{r}\left(\bm{m};\lambda\right)\biggr\},
\end{align}
with $\hat{r}\left(\bm{m};\lambda\right)$ defined in \ref{eq:generalized-partial-tr}
and
\begin{align}
\tilde{\mathcal{J}}\left(\Delta\bm{m};\lambda\right) & =\sum_{\Delta\bm{M}}\tilde{\mathcal{J}}\left(\Delta\bm{m}-\Delta\bm{M},\Delta\bm{M};\lambda\right).\label{eq:kernel-double-prime}
\end{align}
Note that we do not denote the $\tilde{\mathcal{J}}\left(\Delta\bm{m};\lambda\right)$
in a particular manner, and distinguish them from the $\tilde{\mathcal{J}}\left(\delta\bm{M},\Delta\bm{M};\lambda\right)$
by the number of arguments. This shows that only the $\hat{r}\left(\bm{m};\lambda\right)$
are needed in order to calculate the current. That is, we can freely
sum over the $\bm{M}$ and consider only the Cooper pair imbalance
$\bm{m}$, with the information of the CP sector fully encoded into
the $\bm{m}$. 

We can perform a similar simplification of the GME. With Cooper pair
numbers explicitly written, the equation for a term such as $\hat{r}\left(\bm{m},\bm{M};\lambda\right)$
is given by \begin{widetext}
\begin{align}
0 & =(\mathcal{L}_{QD}+\lambda-i\omega_{\bm{m}})\hat{r}\left(\bm{m},\bm{M};\lambda\right)+\sum_{\delta\bm{M}}\sum_{\Delta\bm{M}}\sum_{\bm{m}'}\sum_{\bm{M}'}\bra{\bm{M}+\bm{m}}\left(\hat{S}_{L}^{\dagger}\right)^{\delta M_{L}}\left(\hat{S}_{R}^{\dagger}\right)^{\delta M_{R}}\ket{\bm{M}'+\bm{m}'}\nonumber \\
 & \times\bra{\bm{M}'}\left(\hat{S}_{R}^{\dagger}\right)^{\Delta M_{R}}\left(\hat{S}_{L}^{\dagger}\right)^{\Delta M_{L}}\ket{\bm{M}}\tilde{\mathcal{K}}\left(\delta\bm{M},\Delta\bm{M};\lambda+i\omega_{\bm{m}'}\right)\hat{r}\left(\bm{m}',\bm{M}';\lambda\right),\label{eq:GME-very-expanded}
\end{align}
\end{widetext}where again we have moved the Cooper pairs to the left
and evaluated the $\mathcal{L}_{CP}$ in the same manner as for the
current kernel. Then, we can sum over $\bm{M}$ in \ref{eq:GME-very-expanded}.
Defining 

\begin{align}
\tilde{\mathcal{K}}\left(\Delta\bm{m},\lambda\right)= & \sum_{\Delta\bm{M}}\tilde{\mathcal{K}}\left(\Delta\bm{m}-\Delta\bm{M},\Delta\bm{M};\lambda\right),\label{eq:connection-definitions-kernel}
\end{align}
we obtain the GME as in \ref{eq:GME-generalized-partial-tr}. With
$\Delta\bm{m}=\bm{m}-\bm{m}'$ and $\Delta\bm{M}=\bm{M}'-\bm{M}$,
we see that \ref{eq:connection-definitions-kernel} is equivalent
to \ref{eq:kernel-elements}. In this simplified notation, the term
$\tilde{\mathcal{K}}\left(\Delta\bm{m};\lambda\right)$ includes all
kernel elements that change the number of Cooper pairs by $\Delta\bm{m}$,
regardless of whether the Cooper pair operators act on the right or
on the left. This can be seen from the fact that, in \ref{eq:GME-very-expanded},
$\bm{m}-\bm{m}'=\delta\bm{M}+\Delta\bm{M}.$

From this, the proof of \ref{eq:symmetry-rho} is direct. We can consider
the NZE for a reduced quantum dot operator $\hat{r}\left(\bm{m};i\omega_{\bm{n}}\right)$
\begin{align*}
\left[\mathcal{L}_{QD}-i\omega_{\bm{m}+\bm{n}}\right]\hat{r}\left(\bm{m};i\omega_{\bm{n}}\right)\\
+\sum_{\bm{m}'}\tilde{\mathcal{K}}\left(\bm{m}-\bm{m}',i\omega_{\bm{m}'+\bm{n}}\right)\hat{r}\left(\bm{m}';i\omega_{\bm{n}}\right) & =0.
\end{align*}
This is the same equation as the one for $\hat{r}\left(\bm{m}+\bm{n};0\right)$,
yielding the desired equivalence. Note that this must be satisfied
for all $\bm{m}$, and hence the coefficients relating the two quantities,
namely the $c_{\bm{n}}$, appearing in \ref{eq:symmetry-rho}, must
be independent of $\bm{m}$. 

Within the mean field approximation of our model, these $c_{\bm{n}}$
can be determined by considering the constant of motion 
\begin{equation}
\text{Tr}\{(\hat{S}_{L}^{\dagger})^{n_{L}}(\hat{S}_{R}^{\dagger})^{n_{R}}e^{-i2\bm{n}\cdot\bm{\mu}t/\hbar}\hat{\rho}\left(t\right)\}.\label{eq:constant-of-motion}
\end{equation}
The fact that this quantity is conserved follows from the fact that
the Cooper pair operators commute with the tunneling Hamiltonian.
The time-dependent factor compensates for the time evolution of the
Cooper pair operators due to the chemical potential of the leads.
Since $(\hat{S}_{L}^{\dagger})^{n_{L}}(\hat{S}_{R}^{\dagger})^{n_{R}}$
is purely a system operator, its expectation value can be evaluated
from the reduced density operator, yielding at the initial time
\begin{align*}
\text{Tr}_{S}\{(\hat{S}_{L}^{\dagger})^{n_{L}}(\hat{S}_{R}^{\dagger})^{n_{R}}\hat{\rho}_{S}\left(0\right)\}.
\end{align*}
This is to be compared with its value in a stationary state given
by \ref{eq:steady-state},
\begin{align*}
 & \text{Tr}_{S}\{(\hat{S}_{L}^{\dagger})^{n_{L}}(\hat{S}_{R}^{\dagger})^{n_{R}}\sum_{\bm{n}'}e^{i2\left(\bm{n}'-\bm{n}\right)\cdot\bm{\mu}t/\hbar}\hat{r}_{S}\left(i\omega_{\bm{n}'}\right)\}\\
=\, & \text{Tr}_{QD}\{\sum_{\bm{n}'}e^{i2\left(\bm{n}'-\bm{n}\right)\cdot\bm{\mu}t/\hbar}\hat{r}\left(-\bm{n};i\omega_{\bm{n}'}\right)\}.
\end{align*}
Since $\text{Tr}_{S}\{\hat{r}_{S}\left(\lambda\right)\}=\delta_{\lambda,0}$
due to conservation of probability, together with \ref{eq:symmetry-rho},
we see that $\bm{n}=\bm{n}'$ is the only non-vanishing contribution.
Then, employing this relation again 
\begin{align*}
\text{Tr}_{QD}\{\hat{r}\left(-\bm{n};i\omega_{\bm{n}}\right)\} & =c_{\bm{n}}\text{Tr}_{QD}\{\hat{r}\left(\bm{0};0\right)\}=c_{\bm{n}},
\end{align*}
and comparing this with the initial value, it follows that
\begin{equation}
c_{\bm{n}}=C_{\bm{n}}\left(0\right)=\text{Tr}_{S}\{(\hat{S}_{L}^{\dagger})^{n_{L}}(\hat{S}_{R}^{\dagger})^{n_{R}}\hat{\rho}_{S}\left(0\right)\}.\label{eq:cn-init-time}
\end{equation}
These coefficients can be related to the Fourier components of the
$F\left(\bm{\varphi}\right)$ function of \citep{Siegl2023}. Here,
as mentioned in the main text, we assume that \ref{eq:GME-generalized-partial-tr}
has a single solution for $\lambda=0$. Otherwise, the different solutions
to the GME may not be proportional. 

Let us consider several instances of initial correlations and their
corresponding $c_{\bm{n}}$. First, let us take an initial condition
with no superconducting correlations, so that
\begin{equation}
\hat{\rho}_{S}\left(0\right)=\hat{\rho}_{QD}\left(0\right)\otimes\sum_{\bm{M}}p_{\bm{M}}|\bm{M}\rangle\langle\bm{M}|,\label{eq:trial-state-1}
\end{equation}
for some $\hat{\rho}_{QD}\left(0\right)$, with statistical weights
$p_{\bm{M}}\geq0$ such that $\sum_{\bm{M}}p_{\bm{M}}=1$. The $p_{\bm{M}}$
reflect lack of knowledge over the total number of particles in the
system. In physical terms, a state such as this arises if one considers
that the leads are unconnected at $t=0$, and the tunneling is switched
on adiabatically afterwards, as is usual in the standard Nakajima-Zwanzig
formulation. Then, it follows that $c_{\bm{n}}=\delta_{\bm{n},\bm{0}}$.
Taking into account \ref{eq:current-steady-state-2}, we find as a
result $I^{\infty}\left(t\right)=I_{\bm{0}}$ and no Josephson effect
will develop in the system. 

Next, we consider an initial state with a well defined phase difference
between the two leads, but which still satisfies particle conservation.
To do so, let us introduce the following state in the CP sector
\begin{equation}
\ket{M,\varphi_{0}}=\frac{1}{\sqrt{M+1}}\sum_{m=0}^{M}e^{im\varphi_{0}}\ket{(M-m,m)},
\end{equation}
which has a well-defined total particle number $M=M_{L}+M_{R}$ but
allows for a coherent superposition between states with a different
distribution of Cooper pairs between the two leads. We generalize
this by taking the mixed state
\begin{align}
\hat{\rho}_{S}(0) & =\hat{\rho}_{QD}(0)\otimes\sum_{M=0}^{\infty}p_{M}\ket{M,\varphi_{0}}\bra{M,\varphi_{0}}\label{eq:trial-state-2}\\
 & =\hat{\rho}_{QD}(0)\otimes\sum_{M=0}^{\infty}\sum_{m,m'=0}^{M}\frac{p_{M}}{M+1}\nonumber \\
 & \times e^{i(m-m')\varphi_{0}}\ket{(M-m,m)}\bra{(M-m',m')},\nonumber 
\end{align}
with $\sum_{M}p_{M}=1$. For $p_{M}\geq0$, this corresponds to a
positive semi-definite operator which is moreover Hermitian. Note
that averaging over $\varphi_{0}$ in \ref{eq:trial-state-2} would
yield a state of the form of \ref{eq:trial-state-1}. The coefficients
for this initial state are given by 
\begin{align}
c_{\bm{n}} & =\delta_{n_{L},-n_{R}}e^{i(n_{R}-n_{L})\varphi_{0}/2}\nonumber \\
 & \times\sum_{M=\text{max}\{0,n_{L}\}}^{\infty}p_{M}\frac{M+1-|n_{L}|}{M+1}.
\end{align}
If the statistical weights  $p_{M}$ are centered around a value of
$M$ that is much larger than $|n_{L}|$, the sum is approximately
equal to 1, yielding \ref{eq:particle-cons-init-cond}. This choice
allows us to properly account for relaxation processes in the absence
of a dc-bias,which would bring the phase to the minimum $\varphi_{0}$
of the Josephson potential $V(\varphi)$~\citep{TinkhamBook}.

\section{Analysis of the Cooper pair space structure\label{sec:Relevance-of-the}}

A complication with calculating the current within the theory outlined
above is that the residue operator is infinitely sized, as it incorporates
coherences with infinitely many Cooper pair numbers. However, not
all of these elements contribute in the same manner to the current.
Coherences involving large imbalances in the Cooper pair numbers between
the two leads can only be produced by high order tunneling events.
Hence, in the weak coupling approximation employed here we expect
that they will be negligible. As such, we have not only to consider
which kernel elements are relevant (as was done above) but also which
residue operator coherences are. In this appendix, we develop the
recipe for determining which coherences are relevant at a given perturbation
level. 

\subsection{Relevant contributions to the residue operator\label{subsec:Relevant-contributions-density-op}}

Let us consider first $\hat{r}\left(\bm{0}\right)$. Probability
conservation requires that $\text{Tr}_{QD}\{\hat{r}\left(\bm{0}\right)\}=1$,
therefore the leading term in $\hat{r}\left(\bm{0}\right)$ will be
of order $\Gamma^{0}=1$. 

On the other hand, for a term $\hat{r}\left(\bm{m}\right)$ with $\bm{m}\neq\bm{0}$,
there is no equivalent trace condition. \ref{eq:Liouville-transf-NZE}
gives

\begin{align}
\hat{r}\left(\bm{m}\right)= & \frac{-1}{\mathcal{L}_{QD}-2i\bm{m}\cdot\bm{\mu}/\hbar+\tilde{\mathcal{K}}\left(\bm{0};i2\bm{m}\cdot\bm{\mu}/\hbar\right)}\nonumber \\
\times & \sum_{\bm{m}'\neq\bm{m}}\tilde{\mathcal{K}}\left(\bm{m}-\bm{m}';i2\bm{m}'\cdot\bm{\mu}/\hbar\right)\hat{r}\left(\bm{m}'\right).\label{eq:solution-harmonic}
\end{align}
Hence, in order to obtain the leading order contribution to them we
need to inspect \ref{eq:solution-harmonic} and analyze the relevant
kernels at each order. 

We consider at first only the sequential tunneling approximation
and disregard the denominator in \ref{eq:solution-harmonic}. We know
that $\hat{r}\left(\bm{0}\right)$ is of order 1. Hence, terms such
as $\tilde{\mathcal{K}}_{A}^{\left(1\right)}\left(\pm\bm{u}_{l};\lambda\right)\hat{r}\left(\bm{0}\right)$
will be of order $\Gamma$, since $\tilde{\mathcal{K}}_{A}^{\left(1\right)}\left(\pm\bm{u}_{l};\lambda\right)$
is of order $\Gamma$. This term appears in the sum for $\hat{r}\left(\pm\bm{u}_{l}\right)$,
which will then be of order $\Gamma$ itself. Higher coherences are
not directly connected to $\hat{r}\left(\bm{0}\right).$ For instance,
$\hat{r}[\left(-1,1\right)]$ is connected via $\tilde{\mathcal{K}}_{A}^{\left(1\right)}[\left(1,0\right);\lambda]$
to $\hat{r}[\left(0,1\right)]$, and this is connected to $\hat{r}\left(\bm{0}\right)$
via $\tilde{\mathcal{K}}_{A}^{\left(1\right)}[\left(0,-1\right);\lambda]$.
Hence, $\hat{r}[\left(-1,1\right)]$ will be at least of order $\Gamma^{2}$.
This indicates that the coherences in the residue operator can be
ordered in a \emph{hierarchy}, in which terms which are directly connected
to $\hat{r}\left(\bm{0}\right)$  will be of order $\Gamma$, and
any successive step away will be one order higher in $\Gamma$.

This picture changes significantly when the denominator is taken into
account. The same hierarchy holds in general, but whenever a matrix
element $[\hat{r}\left(\bm{m}\right)]_{\chi'}^{\chi}=\bra{\chi}\hat{r}\left(\bm{m}\right)\ket{\chi'}$
satisfies 
\begin{equation}
|E_{\chi}-E_{\chi'}-2\bm{m}\cdot\bm{\mu}|\simeq\hbar\Gamma,\label{eq:resonance-condition}
\end{equation}
the denominator is of order $\Gamma$ itself and this component and
all of the following components connected to it ``skip'' an order
in this hierarchy. We say then that the coherence is resonant. E.g.
the $\hat{r}\left(\pm\bm{u}_{l}\right)$ are of order 1 at the resonances
of \ref{eq:pair-tunneling-resonance}. 

For the sequential tunneling approximation, provided that the condition
\ref{eq:resonance-condition} is satisfied by at most one value of
$\bm{m}$ (that is, there is no overlap between two or more resonances)
only $\hat{r}\left(\bm{0}\right),\hat{r}(\pm\bm{u}_{l}),\hat{r}[\left(-1,1\right)]$
are relevant for the calculation of the residue operator. Moreover,
$\hat{r}[\left(-1,1\right)]$ is only significant wherever the $\hat{r}\left(\pm\bm{u}_{l}\right)$
are resonant. 

Regarding the second order, the main change in the structure of the
kernel is that now there are also kernel components of the form $\tilde{\mathcal{K}}_{AA}^{\left(2\right)}[\left(-1,1\right);\lambda]$,
$\tilde{\mathcal{K}}_{AA}^{\left(2\right)}[\left(1,-1\right);\lambda]$,
and terms such as $\hat{r}[\left(-1,1\right)$ are now directly coupled
to $\hat{r}\left(\bm{0}\right)$. However, they do so through a kernel
component of order $\Gamma^{2}$ and the hierarchy still holds. The
diagrams involving one normal and one anomalous arc are similar to
the $\tilde{\mathcal{K}}_{A}^{\left(1\right)}\left(\pm\bm{u}_{l};\lambda\right)$
in sequential tunneling but are one order higher in $\Gamma$. As
a result, they are only relevant at resonances such as \ref{eq:pair-tunneling-resonance}
where they give contributions of order $\Gamma^{2}$ to the residue
operator.

Hence, at second order it suffices to consider terms up to $\hat{r}[\left(\pm1,\mp1\right)\pm\bm{u}_{l}]$
to obtain the residue operator at order $\Gamma^{2}$, provided again
that there are no overlapping resonances. Moreover, the $\hat{r}[\left(\pm1,\mp1\right)\pm\bm{u}_{l}]$
are only significant at the pair tunneling resonances. 

\begin{figure}
\begin{centering}
\includegraphics[width=0.85\columnwidth]{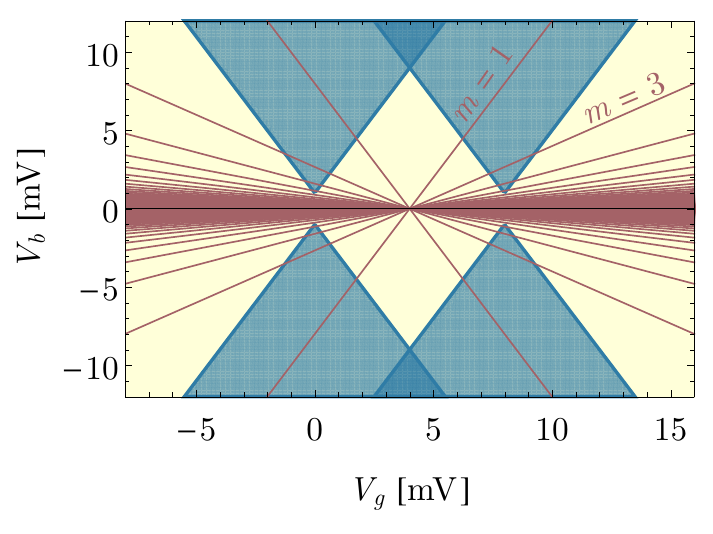}
\par\end{centering}
\caption{Cooper pair resonances in the stability diagram for $U=8,\Delta=0.5$.
Red lines: Cooper pair resonances of increasingly higher order, corresponding
to the condition $2eV_{g}\pm meV_{b}+U=0$.\label{fig:Cooper-pair-resonances}}
\end{figure}

Continuing the logic for higher orders, we can classify the possible
resonances into the two types discussed under \ref{eq:particle-cons-2}.
First, for $\bm{m}=\left(-m,m\right)\pm\bm{u}_{l}$, with $m\in\mathbb{Z}$,
resonances occur at 
\begin{equation}
-2eV_{g}+U\pm meV_{b}=0,
\end{equation}
These are resonant along lines with progressively lower slopes $2/m$
in the stability diagram, until eventually accumulating near $V_{b}=0$
for large $m$. This is shown in \ref{fig:Cooper-pair-resonances}. 

Second, for $\bm{m}=\left(-m,m\right)$, all of the coherences are
resonant only for $V_{b}=0$, regardless of $m$. As such, considering
only the coherences up to a certain $m$ in the residue operator is
justified when sufficiently far ($\sim\Gamma^{k}$) from zero bias.
A resummation can nonetheless be carried out for $V_{b}=0$ exactly~\citep{Governale2008}.
The collapse of the hierarchy at zero bias reflects the degeneracy
of the different Cooper pair number states in that case, since there
is no energy transfer from one lead to the other when a Cooper pair
is exchanged. 

From this, we see that the overlap of resonances in the weak coupling
limit occurs only close to $V_{b}=0$. 

\subsection{Relevant contributions to the current\label{subsec:Relevant-contributions-current}}

We turn now to the current. Therefore, we apply the current kernel
superoperator to the residue operator and inspect the scaling with
$\Gamma$. 

For the sequential tunneling approximation, if no resonances are present,
the main contribution to the dc current will come from $\tilde{\mathcal{J}}_{N}^{\left(1\right)}\left(\bm{0};\lambda\right)\hat{r}\left(\bm{0}\right)$,
which is of order $\Gamma$; terms such as $\tilde{\mathcal{J}}_{A}^{\left(1\right)}(\pm\bm{u}_{l};\lambda)\hat{r}(\mp\bm{u}_{l})$
are of order $\Gamma^{2}$ in general. At the pair-tunneling resonances,
however, the $\tilde{\mathcal{J}}_{A}^{\left(1\right)}(\pm\bm{u}_{l};\lambda)\hat{r}(\mp\bm{u}_{l})$
are of order $\Gamma$ because the $\hat{r}(\pm\bm{u}_{l})$ are of
order 1.  For the Josephson current we find a similar situation.
The only contribution from the first order kernel is $\tilde{\mathcal{J}}_{A}^{\left(1\right)}(\pm\bm{u}_{l};\lambda)\hat{r}(\mp\bm{u}_{\bar{l}})$.
This is of order $\Gamma^{2}$ in general, but of order $\Gamma$
at the CP resonances. 

The second order kernel does not change significantly these estimates.
The main contributions to the Josephson current from the second order
kernel are of order $\Gamma^{2}$ and arise from terms such as $\tilde{\mathcal{J}}_{AA}^{\left(2\right)}[(-1,1);\lambda]\hat{r}\left(\bm{0}\right)$.
The terms originating from kernels with one anomalous and one normal
arc, such as $\tilde{\mathcal{J}}_{NA}^{\left(2\right)}(\pm\bm{u}_{l};\lambda)\hat{r}(\mp\bm{u}_{l'})$,
only contribute significantly at the pair tunneling resonances, where
$\hat{r}(\mp\bm{u}_{l'})$ is of order 1 and hence they give contributions
to the current of order $\Gamma^{2}$. Crucially, these are of order
$\Gamma^{3}$ elsewhere. Hence, proceeding in this manner allows one
to single out all contributions to the current that are at least of
order $\Gamma$ or $\Gamma^{2}$, but also contributions which are
of higher order are included. Another example would be $\tilde{\mathcal{J}}_{N}^{\left(1\right)}\left(\bm{0};\lambda\right)\hat{r}[(-1,1)]$,
which adds to the Josephson current but is $\propto\Gamma^{3}$ for
$eV_{b}\gg\hbar\Gamma$. Alternatively, the secular approximation\emph{~}\citep{Koller2010}
can be employed, in which only terms of order $\Gamma$ and $\Gamma^{2}$
are strictly included; we do not consider it in this work. 

\section{Diagrammatic rules in Liouville space\label{sec:Diagrammatic-rules}}

\begin{figure}
\begin{centering}
\par\end{centering}
\begin{centering}
\includegraphics{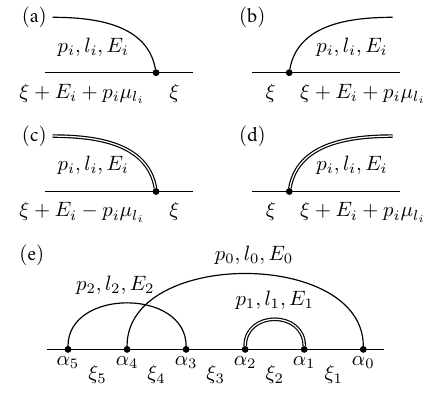}
\par\end{centering}
\caption{Energy conservation at a vertex (\prettyref{rule:cons-E}). (a,c)
For the right vertex of a quasiparticle arc, add $E_{i}+p_{i}\protect\ai_{i}\mu_{l_{i}}$,
to the segment to its left, where $\protect\ai_{i}=+(-)$ for a normal
(anomalous) arc. (b,d) For the left vertex of a quasiparticle arc,
remove $E_{i}+p_{i}\mu_{l_{i}}$ from the segment to its left. $\left(\text{e}\right)$
An exemplary diagram with the energies at every segment written explicitly.
\label{fig:diagrammatic-rules}}

\end{figure}

We summarize the rules needed to translate a given diagram to the
expression $\tilde{\mathcal{K}}_{\cdots\ai_{i}\cdots}^{\left(n\right)}\left(\bm{m},\lambda\right)$.
\begin{enumerate}
\item Draw $2n$ vertices on a \emph{propagation line}. Associate to the
$K$th vertex a Liouville index $\alpha_{K}$. In the following, $K=0$
is the rightmost vertex and $K=2n-1$ is the leftmost. \label{rule:prop}
\item Draw all possible diagrams where each vertex is linked via a \emph{quasiparticle
arc }to another vertex\emph{. }Discard diagrams which can be cut in
two independent diagrams by removing the segment of the propagation
line between two vertices without cutting a quasiparticle arc. The
\emph{i}th quasiparticle arc ($i=1,\ldots,n$) is assigned an energy
$E_{i}$, a lead index $l_{i}$, a Fock index $p_{i}$, a spin $\sigma_{i}$
and an index $\ai_{i}$ which can be either normal (represented by
a single arc) or anomalous (double arc). \label{rule:labels}
\item Starting from the rightmost vertex and following the propagation line
until the leftmost vertex: \label{rule:multip}
\begin{enumerate}
\item For the $K$th vertex, if it is the left vertex of the $i$th quasiparticle
arc, multiply by $\left(\alpha_{K}\right)^{K}\fsop_{\sigma_{i}}^{\bar{p}_{i}\alpha_{K}}$;
if it is the right vertex of a normal arc, multiply by $\left(\alpha_{K}\right)^{K+1}\fsop_{\sigma_{i}}^{p_{i}\alpha_{K}}$;
if it is the right vertex of an anomalous arc, multiply by $\left(\alpha_{K}\right)^{K+1}\fsop_{\bar{\sigma}_{i}}^{\bar{p}_{i}\alpha_{K}}$.
\label{rule:vertices}
\item For the $J$th segment of the propagation line between two vertices
($J=1,\ldots,2n-1$), multiply by the propagator
\[
\frac{1}{\xi_{J}-i\hbar\mathcal{L}_{QD}+i\hbar\lambda}.
\]
The $\xi_{J}$ of the segment can be determined by conservation of
energy in each vertex. Given two segments separated by the \emph{right}
vertex of a quasiparticle arc, the energy of the segment to the left
must be larger by $E_{i}+p_{i}\ai_{i}\mu_{i}$, where $\ai_{i}=+(-)$
for a normal (anomalous) arc. Given two segments separated by the
\emph{left} vertex of an arc, the energy of the segment to the right
must be larger by $E_{i}+p_{i}\mu_{i}$. This is shown  in \ref{fig:diagrammatic-rules}~$\left(\text{a}\right)-\left(\text{d}\right)$.
\label{rule:cons-E}
\end{enumerate}
\item For each quasiparticle arc, multiply by $|t_{l_{i}}|^{2}f\left(\alpha_{K}E_{i}\right)$,
with $\alpha_{K}$ the index of its rightmost vertex. If the arc is
normal, multiply by $\dos_{N,l_{i}}\left(E_{i}\right)$; if it is
anomalous, multiply by $p_{i}\sigma_{i}e^{-ip_{i}\varphi_{l_{i}}}\dos_{A,l_{i}}\left(E_{i}\right)\text{sgn}\left(E_{i}\right)$.
Multiply by $-1$ for each crossing of two quasiparticle arcs. \label{rule:fermi-DOS}
\item Sum over the internal (spin) and superoperator indices and integrate
over the energies $E_{i}$ and multiply by $1/i\hbar$. I.e.\label{rule:sum}
\[
\frac{1}{i\hbar}\sum_{\left\{ \alpha_{K}\right\} }\sum_{\left\{ \sigma_{i}\right\} }\int\cdots dE_{i}\cdots.
\]
For each combination of $\{\psi_{i}\},\{l_{i}\},\{p_{i}\}$, the corresponding
diagram is assigned to the kernel $\tilde{\mathcal{K}}_{\cdots\ai_{i}\cdots}^{\left(n\right)}\left(\sum_{i}\delta_{\psi_{i}A}p_{i}\bm{u}_{l_{i}},\lambda\right)$.
\item To obtain the contribution to the current kernel, multiply by $ep_{n}\delta_{\alpha_{2n-1},+}\ell_{n}/2$,
where $n$ is the leftmost quasiparticle arc and $\ell_{n}=\left(-1\right)^{\delta_{l_{n},L}}$.\label{rule:current}
\end{enumerate}
\prettyref{rule:current}  follows from the expression of the symmetrized
current operator and the tunneling Liouvillian
\begin{align}
\mathcal{L}_{T} & =\frac{1}{i\hbar}\sum_{lp\bm{k}\sigma\alpha}pt_{lk}^{p}\mathcal{C}_{l\sigma\bm{k}}^{p\alpha}\fsop_{\sigma}^{\bar{p}\alpha},\\
\hat{I} & =\frac{e}{2i\hbar}\sum_{lp\bm{k}\sigma}\ell t_{lk}^{p}\mathcal{C}_{l\sigma\bm{k}}^{p+}\fsop_{\sigma}^{\bar{p}+},
\end{align}
where $\mathcal{C}_{l\sigma k}^{p+}\bullet=\hat{c}_{l\sigma k}^{p}\bullet$
and $\mathcal{C}_{l\sigma k}^{p-}\bullet=\bullet\hat{c}_{l\sigma k}^{p}$. 

As an example, let us give the expression for the 3rd order diagram
represented in \ref{fig:diagrammatic-rules}~$\left(\text{e}\right)$.
We first obtain the lead energies in each segment of the propagator
line, labeled $\xi_{i},$ $i=1-5$ in the diagram. Using \prettyref{rule:cons-E}
we obtain
\begin{align}
\xi_{1} & =E_{0}+p_{0}\mu_{l_{0}},\\
\xi_{2} & =E_{0}+E_{1}+p_{0}\mu_{l_{0}}-p_{1}\mu_{l_{1}},\\
\xi_{3} & =E_{0}+p_{0}\mu_{l_{0}}-2p_{1}\mu_{l_{1}},\\
\xi_{4} & =E_{0}+E_{2}+p_{0}\mu_{l_{0}}-2p_{1}\mu_{l_{1}}+p_{2}\mu_{l_{2}},\\
\xi_{5} & =E_{2}-2p_{1}\mu_{l_{1}}+p_{2}\mu_{l_{2}}.
\end{align}
This yields the following contribution 
\begin{align*}
 & \frac{-1}{i\hbar}\sum_{\left\{ \alpha_{K}\right\} _{K=0}^{5}}\sum_{\left\{ \sigma_{i}\right\} _{i=0}^{2}}\int dE_{2}\int dE_{1}\int dE_{0}\\
\times & \alpha_{2}\alpha_{0}\sigma_{1}p_{1}\fsop_{\sigma_{2}}^{\bar{p}_{2}\alpha_{5}}\\
\times & \frac{1}{E_{2}+p_{2}\mu_{l_{2}}-2p_{1}\mu_{l_{1}}-i\hbar\mathcal{L}_{S}+i\hbar\lambda}\fsop_{\sigma_{0}}^{\bar{p}_{0}\alpha_{4}}\\
\times & \frac{|t_{l_{2}}|^{2}f\left(\alpha_{3}E_{2}\right)\dos_{N,l_{2}}\left(E_{2}\right)}{E_{2}+p_{2}\mu_{l_{2}}+E_{0}+p_{0}\mu_{l_{0}}-2p_{1}\mu_{l_{1}}-i\hbar\mathcal{L}_{QD}+i\hbar\lambda}\fsop_{\sigma_{2}}^{p_{2}\alpha_{3}}\\
\times & \frac{1}{E_{0}+p_{0}\mu_{l_{0}}-2p_{1}\mu_{l_{1}}-i\hbar\mathcal{L}_{QD}+i\hbar\lambda}\fsop_{\sigma_{1}}^{\bar{p}_{1}\alpha_{2}}\\
\times & \frac{|t_{l_{1}}|^{2}f\left(\alpha_{1}E_{1}\right)e^{-ip_{1}\varphi_{l_{1}}}\dos_{A,l_{1}}\left(E_{1}\right)\text{sgn}\left(E_{1}\right)}{E_{0}+p_{0}\mu_{l_{0}}+E_{1}-p_{1}\mu_{l_{1}}-i\hbar\mathcal{L}_{QD}+i\hbar\lambda}\fsop_{\bar{\sigma}_{1}}^{\bar{p}_{1}\alpha_{1}}\\
\times & \frac{|t_{l_{0}}|^{2}f\left(\alpha_{0}E_{0}\right)\dos_{N,l_{0}}\left(E_{0}\right)}{E_{0}+p_{0}\mu_{l_{0}}-i\hbar\mathcal{L}_{QD}+i\hbar\lambda}\fsop_{\sigma_{0}}^{p_{0}\alpha_{0}}.
\end{align*}
From \prettyref{rule:labels} we see that this belongs to $\tilde{\mathcal{K}}_{NAN}^{\left(3\right)}\left(p_{1}\bm{u}_{l_{1}},\lambda\right)$.
The corresponding diagram for the current kernel can be obtained by
taking and multiplying by $ep_{2}\ell_{2}\delta_{\alpha_{5},+}/2$
within the summations. 

\section{First order integrals\label{sec:seq-tun-integrals-1}}

\begin{figure}
\begin{centering}
\par\end{centering}
\begin{centering}
\includegraphics{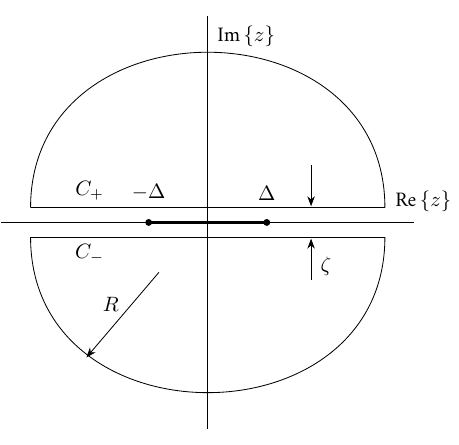}
\par\end{centering}
\caption[Contours $\Gamma_{\pm}$ for the integration of the sequential
tunneling integral $I_{\text{N},l}^{q}\left(\nu\right)$.]{ Contours
$C_{\pm}$ for the integration of the sequential tunneling integral
$I_{\text{N},l}^{q}\left(\nu\right)$. The branch cut of $g_{l}\left(z\right)$
at $\left[-\left|\Delta_{l}\right|,\left|\Delta_{l}\right|\right]$
are marked by thick lines. \label{fig:seq-tun-contour}}
\end{figure}

The sequential tunneling integral is given by \ref{eq:kernel-2nd-order}.
It can be calculated employing the residue theorem
\begin{align}
I_{N,l}^{q}\left(\nu\right) & =\int_{-\infty}^{\infty}dE\frac{f\left(qE\right)\dos_{N,l}\left(E\right)}{E-\nu+i0^{+}}\nonumber \\
 & =\lim_{W\to\infty}\int_{-\infty}^{\infty}dE\frac{f\left(qE\right)\dos_{N,l}\left(E\right)L\left(E,W\right)}{E-\nu+i0^{+}}.\label{eq:int-seq-tun}
\end{align}
Here, a Lorentzian cutoff $L\left(E,W\right)=W^{2}/(E^{2}+W^{2})$
has been introduced to regularize the integral for large $E$. We
introduce also the function 
\begin{equation}
g_{N,l}\left(E\right)=\dos_{l.0}\sqrt{\frac{E^{2}}{E^{2}-|\Delta_{l}|^{2}}},
\end{equation}
such that $\dos_{N,l}\left(E\right)=\real\{g_{N,l}\left(E\right)\}$.
We define the analytic continuation of $g_{N,l}\left(E\right)$ having
a branch cut at $\real\left\{ z\right\} \in\left[-\left|\Delta_{l}\right|,\left|\Delta_{l}\right|\right]$.
In order to calculate $I_{N,l}^{q}\left(\nu\right)$, we define the
contour $C_{\pm}$ (sketched in \ref{fig:seq-tun-contour}) as a semicircle
extending in the upper (lower) half-plane with radius $R$ and separated
by a distance $\zeta$ from the real axis. In this manner, we have
\begin{align}
 & \int_{-\infty}^{\infty}dE\frac{f\left(qE\right)\dos_{N,l}\left(E\right)L\left(E,W\right)}{E-\nu+i0^{+}}\nonumber \\
= & \lim_{R\to\infty}\lim_{\zeta\to0^{+}}\sum_{s=\pm}\frac{s}{2}\int_{C_{s}}dE\frac{f\left(qE\right)g_{N,l}\left(E\right)L\left(E,W\right)}{E-\nu+i0^{+}}.
\end{align}
The integrals over $C_{s}$ can be calculated employing the residue
theorem. The Fermi function $f(E)$ has poles at $iE_{k}=i\hbar\omega_{k}$
with residue $-\beta^{-1}$, where $\omega_{k}=2\pi i\left(k+1/2\right)/\beta\hbar$,
$k\in\mathbb{Z}$ are the Matsubara frequencies. The Lorentzian has
poles at $E=\pm iW$ with residue $\left(1/2\right)\left(\mp iW\right)$.
Taking the limits, the integral yields
\begin{align}
I_{N,l}^{q}\left(\nu\right) & =-i\pi f\left(q\nu\right)g_{N,l}\left(\nu-i0^{+}\right)\label{eq:seq-tun-int}\\
 & -q\left[\mathcal{S}_{N,l}^{\left(2\right)}\left(\nu\right)-\lim_{W\to\infty}\mathcal{C}_{N,l}^{\left(2\right)}\left(W\right)\right],\nonumber \\
S_{N,l}^{\left(2\right)}\left(\nu\right) & =\frac{2\pi}{\beta}\sum_{k=0}^{\infty}\frac{E_{k}g_{N,l}\left(iE_{k}\right)}{E_{k}^{2}+\nu^{2}},\label{eq:Matsubara-sum-2}\\
C_{N,l}^{\left(2\right)}\left(W\right) & =\frac{2\pi}{\beta}\sum_{k=0}^{\infty}\frac{1}{2}\left[\frac{g_{N,l}\left(iE_{k}\right)+g_{N,l}\left(iW\right)}{E_{k}+W}\right]\nonumber \\
 & +\frac{2\pi}{\beta}\sum_{k=0}^{\infty}\frac{1}{2}\left[\frac{g_{N,l}\left(iE_{k}\right)-g_{N,l}\left(iW\right)}{E_{k}-W}\right].\label{eq:Divergence-constant}
\end{align}
The infinitesimal factor $i0^{+}$ reflects that the function is evaluated
below the branch cut. This turns out to be only relevant in the calculation
of the imaginary part of $g_{N,l}\left(\nu-i0^{+}\right)$. The integral
satisfies the property $I_{N,l}^{q}\left(\nu\right)=-[I_{N,l}^{\bar{q}}\left(-\nu\right)]^{*}$,
which guarantees the \hbox{hermiticity} of the density operator.

The real valued function $g_{N,l}\left(iE\right)$ is mostly flat
except for the region $\left|E\right|\lesssim|\Delta_{l}|$, where
it exhibits a dip. For $\beta|\Delta_{l}|\ll1$, the sampling at the
Matsubara frequencies ignores the dip and the density of states is
effectively flat. In that case, we recover the formula for non-superconducting
leads with a flat density of states~\citep{KollerThesis}
\begin{align}
 & \lim_{\beta\left|\Delta_{l}\right|\to0}S_{N,l}^{\left(2\right)}\left(\nu\right)=\\
 & -\dos_{0,l}\left[\real\left\{ \Psi^{\left(0\right)}\left(\frac{1}{2}+\frac{i\beta\nu}{2\pi}\right)\right\} +\gamma_{\Psi}\right],\nonumber \\
 & \lim_{\beta\left|\Delta_{l}\right|\to0}C_{N,l}^{\left(2\right)}\left(W\right)=\\
 & -\dos_{0,l}\left[\real\left\{ \Psi^{\left(0\right)}\left(\frac{1}{2}+\frac{\beta W}{2\pi}\right)\right\} +\gamma_{\Psi}\right],\nonumber 
\end{align}
where $\gamma_{\Psi}=\sum_{k=1}^{\infty}\log\left(1+1/k\right)$.
Note that the $C_{N,l}^{\left(2\right)}\left(W\right)\sim\log\beta W$
encapsulates the logarithmic divergence in the limit $\beta W\to\infty$.

In the opposite (low temperature) limit, $\beta\left|\Delta_{l}\right|\gg1$,
the sampling of $g_{N,l}\left(iE\right)$ by the Matsubara energies
$E_{k}$ is taken as such small intervals that the sum over $E_{k}$
can be converted to an integral, with the differential defined as
$\lim_{\beta\left|\Delta_{l}\right|\to\infty}2\pi/\beta\left|\Delta_{l}\right|$.
This limit can be performed most conveniently by taking a hard cutoff
$\Theta\left(W^{2}-E^{2}\right)$ instead of the Lorentzian in \ref{eq:int-seq-tun}.
Then, employing again the residue theorem and converting the resulting
sums to integrals, one obtains
\begin{equation}
\lim_{\beta\left|\Delta_{l}\right|\to\infty}S_{N,l}^{\left(2\right)}\left(\nu\right)=\frac{\dos_{0,l}}{2}\log\frac{\nu^{2}}{\nu^{2}+W^{2}}.
\end{equation}

The equivalent of \ref{eq:int-seq-tun} for the anomalous kernel is
given by
\begin{equation}
I_{A,l}^{q}\left(\nu\right)=\int_{-\infty}^{\infty}dE\frac{f\left(qE\right)\dos_{A,l}\left(E\right)\text{sgn}\left(E\right)}{E-\nu+i0^{+}}.
\end{equation}
Contrary to the normal case, no Lorentzian regularization is needed,
since the anomalous DOS decays as $\sim E^{-1}$ for large $E$. 

Before proceeding to contour integration, a question arises as to
the analytic extension of the $\text{sgn\ensuremath{\left(E\right)} }$function.
Let us first define the function 
\begin{equation}
g_{A,l}\left(E\right)=\dos_{0,l}\sqrt{\frac{|\Delta_{l}|^{2}}{E^{2}-|\Delta_{l}|^{2}}}.
\end{equation}
If we choose the branch cut in its analytic continuation to lie in
$\real\left\{ z\right\} \in\left[-\left|\Delta_{l}\right|,\left|\Delta_{l}\right|\right]$,
$g_{A,l}\left(E\right)$ has an extra branch cut on the imaginary
axis. Fortunately, the branches can be chosen to remove this spurious
branch cut. This is equivalent to considering the analytic continuation
of $\text{sgn}\left(E\right)$ as $\text{sgn}\left(\real\left\{ E\right\} \right)$,
which is the natural choice to take. With this, we can proceed as
above to find 

\begin{align}
I_{A,l}^{q}\left(\nu\right) & =-i\pi f\left(q\nu\right)g_{A,l}\left(\nu-i0^{+}\right)\text{sgn}\left(\nu\right)-q\mathcal{S}_{A,l}^{\left(2\right)}\left(\nu\right),\label{eq:seq-tun-anom-int}\\
S_{A,l}^{\left(2\right)}\left(\nu\right) & =\frac{2\pi}{\beta}\sum_{k=0}^{\infty}\frac{i\nu g_{A,l}\left(iE_{k}\right)}{E_{k}^{2}+\nu^{2}}.\label{eq:anomalous-matsubara-S2}
\end{align}
Note that the function $S_{A,l}^{\left(2\right)}\left(\nu\right)$
is real, since $g_{A,l}\left(i\nu\right)$ is purely imaginary. The
anomalous second order integral satisfies $I_{A,l}^{q}\left(\nu\right)=[I_{A,l}^{\bar{q}}\left(-\nu\right)]^{*}$.
Since $g_{A,l}\left(x\right)\simeq\dos_{0,l}|\Delta_{l}|/x$ for $x\gg|\Delta_{l}|$,
the Matsubara sum vanishes in the limit $\beta|\Delta_{l}|\ll1$ .
In the opposite limit $\beta|\Delta_{l}|\gg1$, we can transform again
the sum into an integral and solve it analytically, yielding
\begin{align}
\lim_{\beta|\Delta_{l}|\to\infty}S_{A,l}^{\left(2\right)}\left(\nu\right) & =g_{A,l}\left(\nu\right)\text{sgn}\left(\nu\right)\\
 & \times\text{arctanh}\left[\lim_{E/|\Delta_{l}|\to\infty}\frac{g_{N,l}\left(iE\right)}{g_{N,l}\left(\nu\right)}\right],\nonumber 
\end{align}

\section{Second order kernel and integrals\label{sec:Cotunneling-integrals}}

We discuss in detail the second order kernels and discuss the numerical
implementation employed to obtain the current and the density operator.
We use the notation $\tilde{\mathcal{K}}_{\ai'\ai}^{\left(2x\right)}\left(\bm{m};\lambda\right)$,
with $x\in\{\text{D,\,X}\}$ denoting the class of the diagram. Let
us consider first the fully normal kernel ($\ai'\ai=NN$). After summing
over the lead and Fock indices of the two lines, the D contribution
is given by 
\begin{align}
\tilde{\mathcal{K}}_{NN}^{\left(2\text{D}\right)}\left(\bm{0};\lambda\right)= & \frac{1}{i\hbar}\sum_{l'l}\sum_{\sigma'\sigma}\sum_{p'p}\sum_{\left\{ \alpha_{i}\right\} }\alpha_{3}\alpha_{0}\nonumber \\
\times & \left|t_{l'}\right|^{2}\left|t_{l}\right|^{2}\int_{-\infty}^{\infty}dE'\int_{-\infty}^{\infty}dE\fsop_{\sigma}^{\bar{p}\alpha_{3}}\nonumber \\
\times & \frac{\dos_{N,l'}\left(E'\right)f\left(\alpha_{1}E'\right)}{E-i\hbar\mathcal{L}_{QD}+p\mu_{l}+i\hbar\lambda}\fsop_{\sigma'}^{\bar{p}'\alpha_{2}}\nonumber \\
\times & \frac{1}{E+E'-i\hbar\mathcal{L}_{QD}+p\mu_{l}+p'\mu_{l'}+i\hbar\lambda}\fsop_{\sigma'}^{p'\alpha_{1}}\nonumber \\
\times & \frac{\dos_{N,l}\left(E\right)f\left(\alpha_{0}E\right)}{E-i\hbar\mathcal{L}_{QD}+p\mu_{l}+i\hbar\lambda}\fsop_{\sigma}^{p\alpha_{0}}.\label{eq:K4D-complete}
\end{align}
Meanwhile, the fully normal X kernel is given by 
\begin{align}
\tilde{\mathcal{K}}_{NN}^{\left(2\text{X}\right)}\left(\bm{0};\lambda\right)= & \frac{-1}{i\hbar}\sum_{l'l}\sum_{\sigma'\sigma}\sum_{p'p}\sum_{\left\{ \alpha_{i}\right\} }\alpha_{3}\alpha_{0}\nonumber \\
\times & \left|t_{l'}\right|^{2}\left|t_{l}\right|^{2}\int_{-\infty}^{\infty}dE'\int_{-\infty}^{\infty}dE\fsop_{\sigma'}^{\bar{p}'\alpha_{3}}\nonumber \\
\times & \frac{\dos_{N,l'}\left(E'\right)f\left(\alpha_{1}E'\right)}{E'-i\hbar\mathcal{L}_{QD}+p'\mu_{l'}+i\hbar\lambda}\fsop_{\sigma}^{\bar{p}\alpha_{2}}\nonumber \\
\times & \frac{1}{E+E'-i\hbar\mathcal{L}_{QD}+p\mu_{l}+p'\mu_{l'}+i\hbar\lambda}\fsop_{\sigma'}^{p'\alpha_{1}}\nonumber \\
\times & \frac{\dos_{N,l}\left(E\right)f\left(\alpha_{0}E\right)}{E-i\hbar\mathcal{L}_{QD}+p\mu_{l}+i\hbar\lambda}\fsop_{\sigma}^{p\alpha_{0}}.\label{eq:K4X-complete}
\end{align}
For the mixed and anomalous contributions, it suffices to apply the
following rules for each normal line that is turned from normal into
anomalous, as per the diagrammatic rules. First, the following substitutions
are performed:
\begin{align}
\fsop_{\sigma_{i}}^{p_{i}\alpha_{K}} & \to\fsop_{\bar{\sigma}_{i}}^{\bar{p}_{i}\alpha_{K}},\\
\dos_{N,l_{i}}\left(E_{i}\right) & \to p_{i}\sigma_{i}e^{-ip_{i}\varphi_{l_{i}}}\dos_{A,l_{i}}\left(E_{i}\right)\text{sgn}\left(E_{i}\right),
\end{align}
where the first refers to the rightmost vertex of the quasiparticle
line. Then, for each propagator to the left of this vertex, we have
to further substitute:
\begin{equation}
i\hbar\mathcal{L}_{QD}\to i\hbar\mathcal{L}_{QD}+2p_{i}\mu_{l_{i}}/\hbar,
\end{equation}
in order to account for \prettyref{rule:cons-E} of the diagrammatic
rules. Finally, the term now contributes to the kernel $\tilde{\mathcal{K}}(\bm{m}+p\bm{u}_{l'};\lambda)$.
For instance, let us turn the two lines of \ref{eq:K4D-complete}
from normal to anomalous yields. Now, the terms corresponding for
the different values of $p,l,l'$ will now contribute to distinct
kernel components. Using the rules, we find 
\begin{align}
 & \tilde{\mathcal{K}}_{AA}^{\left(2\text{D}\right)}[p\left(\bm{u}_{l}-\bm{u}_{l'}\right);\lambda]\nonumber \\
= & \frac{-1}{i\hbar}\sum_{\sigma'\sigma}\sum_{\left\{ \alpha_{i}\right\} }\sigma'\sigma\alpha_{3}\alpha_{0}e^{ip\left(\varphi_{l'}-\varphi_{l}\right)}\nonumber \\
\times & \left|t_{l'}\right|^{2}\left|t_{l}\right|^{2}\int_{-\infty}^{\infty}dE'\int_{-\infty}^{\infty}dE\fsop_{\sigma}^{\bar{p}\alpha_{3}}\nonumber \\
\times & \frac{\dos_{A,l'}\left(E'\right)\text{sgn}\left(E'\right)f\left(\alpha_{1}E'\right)}{E-i\hbar\mathcal{L}_{QD}-p\mu_{l}+2p\mu_{l'}+i\hbar\lambda}\fsop_{\sigma'}^{p\alpha_{2}}\nonumber \\
\times & \frac{1}{E+E'-i\hbar\mathcal{L}_{QD}-p\mu_{l}+p\mu_{l'}+i\hbar\lambda}\fsop_{\bar{\sigma}'}^{p\alpha_{1}}\nonumber \\
\times & \frac{\dos_{A,l}\left(E\right)\text{sgn}\left(E\right)f\left(\alpha_{0}E\right)}{E-i\hbar\mathcal{L}_{QD}-p\mu_{l}+i\hbar\lambda}\fsop_{\bar{\sigma}}^{\bar{p}\alpha_{0}}.\label{eq:anomalous D kernel}
\end{align}
Here we have already taken into account that the only terms that are
not zero for the QD Josephson junction have $p'=\bar{p}$. The corresponding
current kernels are obtained by taking $\alpha_{3}=+$ and multiplying
by $ep'\ell/2$.

The integrals that need to be calculated for these kernels have the
form

\begin{align}
I_{\ai'\ai,l'l}^{\text{D},q'q}\left(\nu,\nu',\Lambda\right) & =\lim_{\eta\to0^{+}}\int_{-\infty}^{\infty}dE'\int_{-\infty}^{\infty}dE\times\label{eq:cotunneling-integral-D-1-1}\\
\frac{f\left(q'E'\right)\dos_{\ai',l'}\left(E^{\prime}\right)}{E-\nu'+i\eta} & \frac{s_{\psi}\left(E\right)s_{\psi'}\left(E'\right)}{E+E'-\Lambda+i\eta}\frac{f\left(qE\right)\dos_{\ai,l}\left(E\right)}{E-\nu+i\eta},\nonumber \\
I_{\ai'\ai,l'l}^{\text{X},q'q}\left(\nu,\nu',\Lambda\right) & =\lim_{\eta\to0^{+}}\int_{-\infty}^{\infty}dE'\int_{-\infty}^{\infty}dE\times\label{eq:cotunneling-integral-X-1-1}\\
\frac{f\left(q'E'\right)\dos_{\ai',l'}\left(E^{\prime}\right)}{E'-\nu'+i\eta} & \frac{s_{\psi}\left(E\right)s_{\psi'}\left(E'\right)}{E+E'-\Lambda+i\eta}\frac{f\left(qE\right)\dos_{\ai,l}\left(E\right)}{E-\nu+i\eta},\nonumber 
\end{align}
where 
\begin{equation}
s_{\psi}\left(E\right)=\begin{cases}
\text{sgn}\left(E\right) & \psi=A,\\
1 & \psi=N.
\end{cases}
\end{equation}
Employing partial fraction decomposition, these integrals can be
written as

\begin{align}
I_{\ai'\ai,l'l}^{\text{D},q'q}\left(\nu,\nu',\Lambda\right) & =\frac{J_{\ai'\ai,l'l}^{q'q}\left(\nu,\Lambda\right)-J_{\ai'\ai,l'l}^{q'q}\left(\nu',\Lambda\right)}{\nu-\nu'},\label{eq:ID-func-D}\\
I_{\ai'\ai,l'l}^{\text{X},q'q}\left(\nu,\nu',\Lambda\right) & =I_{\ai',l'}^{q'}\left(\nu'\right)\frac{I_{\ai,l}^{q}\left(\nu\right)-I_{\ai,l}^{q}\left(\Lambda-\nu'\right)}{\nu+\nu'-\Lambda}\nonumber \\
 & -\frac{J_{\ai'\ai,l'l}^{q'q}\left(\nu,\Lambda\right)-J_{\ai'\ai,l'l}^{q'q}\left(\Lambda-\nu',\Lambda\right)}{\nu+\nu'-\Lambda},\label{eq:ID-func-X}
\end{align}
where we have defined the function

\begin{align}
 & J_{\ai'\ai,l'l}^{q'q}\left(\nu,\Lambda\right)=\nonumber \\
 & \lim_{\eta\to0^{+}}\int_{-\infty}^{\infty}dE'\int_{-\infty}^{\infty}dEs_{\psi'}\left(E'\right)s_{\psi}\left(E\right)\nonumber \\
 & \times\frac{f\left(q'E'\right)\dos_{\ai',l'}\left(E'\right)}{E-\nu+i\eta}\frac{f\left(qE\right)\dos_{\ai,l}\left(E\right)}{E+E'-\Lambda+i\eta},\label{eq:J-integral}
\end{align}
on which we will focus. In the limits $\nu\to\nu'$ and $\Lambda\to\nu+\nu'$,
the integrals of \ref{eq:ID-func-D,eq:ID-func-X} are equal to $\partial_{\nu}J_{\ai'\ai,l'l}^{q'q}\left(\nu,\Lambda\right)$
and $I_{\ai',l'}^{q'}\left(\Lambda-\nu\right)\partial_{\nu}I_{\ai,l}^{q}\left(\nu\right)-\partial_{\nu}J_{\ai'\ai,l'l}^{q'q}\left(\nu,\Lambda\right)$,
respectively. 

\begin{figure}
\begin{centering}
\includegraphics[scale=0.75]{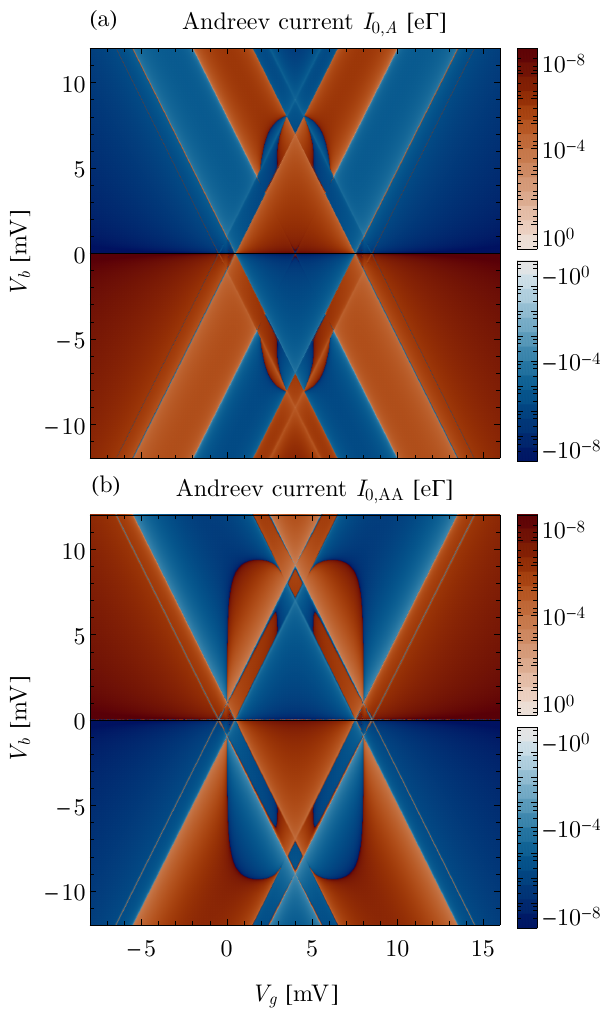}
\par\end{centering}
\caption{Andreev current components. (a) Andreev current $I_{0,A}$ originating
from the first order kernel as a function of the gate and bias voltages,
$V_{g}$ and $V_{b}$, respectively, in logarithmic scale for the
same parameters as in \ref{fig:Andreev-current}. (b) Andreev current
$I_{0,AA}$ originating from the second order kernel. \label{fig:Andreev-1}}
\end{figure}

In order to obtain the results presented in this work, we calculated
the integral numerically on a grid and then obtained the derivative
through finite differences. Small numerical errors appears when
sampling the coherence peaks of the density of states. To help with
convergence of the integrals, we consider $\dos_{\ai,l}\left(E\right)\to\real\{g_{\ai,l}\left(E-i\zeta\right)\}$
in the grids, and set the Dynes parameter $\zeta$ accordingly. 

Lastly, employing the symmetry properties of this function, in particular
\begin{equation}
J_{\ai'\ai,l'l}^{q'q}\left(\nu,\Lambda\right)=\psi\psi'[J_{\ai'\ai,l'l}^{\bar{q}'\bar{q}}\left(-\nu,-\Lambda\right)]^{*},
\end{equation}
with the shorthand notation $\psi\equiv\left(-1\right)^{\delta_{\psi,A}}$,
it is possible to reduce the number of calculated grids by half.
Even then, eight grids need to be calculated in the case of identical
gaps for both leads, and four times that number for different gaps.
Fortunately, \ref{eq:J-integral} can be reduced to one-dimensional
integrals by solving for the variable $E'$ using the expression for
the sequential tunneling integral given above. 

\section{Contributions to the Andreev current\label{sec:Contributions-andreev}}

We discuss separately the two contributions to the Andreev current,
$I_{0,A}$ and $I_{0,AA}$, which are represented in \ref{fig:Andreev-1}.
We first point out that the two terms cancel inside the Coulomb diamonds
almost exactly. This signals that the Andreev current is, outside
of the resonances, intrinsically a second order process and that these
two contributions must be understood together. We say here ``almost
exactly'' because higher order contributions such as those discussed
in \ref{subsec:Andreev-current} result in small differences. This
can be appreciated here for the pair tunneling resonances close to
$eV_{b}=0,eV_{g}=4\,\mV$, where $I_{0,A}$ shows a change in the
current that is not observed in $I_{0,AA}$. 

For $I_{0,A}$ we observe that the current changes sign close to but
not exactly at the resonances, where the current is actually finite,
as can be seen in \ref{fig:Andreev-current}~(c). For $I_{0,AA}$,
we see the vertical features for $eV_{g}=0,U$, which appeared as
dips in the differential conductance in \ref{fig:Andreev-current}.
For $eV_{g}=0$ we can study this feature employing an infinite interaction
approximation to the Andreev current. Provided that $\mu_{l}\ll U$,
we find 
\begin{align}
 & \lim_{U\to\infty}I_{0,AA}=2e\sum_{l}\ell\Gamma_{l}^{2}\frac{1}{\mu_{l}}\frac{\hbar}{2\pi}\\
\times & \dos_{A,l}\left(\mu_{l}\right)\text{sgn}\left(\mu_{l}\right)S_{A,l}^{\left(2\right)}\left(\mu_{l}\right)\left[f\left(\mu_{l}\right)p_{1}/2-f\left(-\mu_{l}\right)p_{0}\right],\nonumber 
\end{align}
where $p_{0}=\bra{0}\hat{r}\left(\bm{0}\right)\ket{0}$ is the population
of the unoccupied state and $p_{1}=\sum_{\sigma}p_{\sigma}$ with
$p_{\sigma}=\bra{\sigma}\hat{r}\left(\bm{0}\right)\ket{\sigma}$.
For $eV_{b}>\Delta$, along the vertical line $eV_{g}=0$ in the stability
diagram $p_{1}=2p_{0}$ due to symmetry, and we find 
\begin{align*}
 & \lim_{U\to\infty}I_{0,AA}=-2e\sum_{l}\ell\Gamma_{l}^{2}\frac{1}{2\ell eV_{b}}\frac{\hbar}{2\pi}\\
 & \times\dos_{A,l}\left(\ell eV_{b}\right)\text{sgn}\left(\ell eV_{b}\right)S_{A,l}^{\left(2\right)}\left(\ell eV_{b}\right)\tanh\left(\ell\beta eV_{b}/2\right).
\end{align*}
Since $S_{A,l}^{\left(2\right)}\left(\nu\right)$ is odd in $\nu$,
{[}see \ref{eq:anomalous-matsubara-S2}{]}, for equal tunneling rates
$\Gamma_{l}\equiv\Gamma$ and absolute values of the gap $|\Delta_{l}|\equiv\Delta$,
the sum over the two leads cancels and the Andreev current is zero.
This can be seen clearly in \ref{fig:Andreev-1}~(b) for $\Delta<eV_{b}\ll U$.
For $eV_{g}=U$, the \hbox{analysis} is the same due to particle-hole
symmetry around $eV_{g}=U/2$. 

We also remark the absence of features along the pair tunneling resonances
for $I_{0,AA}$, which as pointed out above originate exclusively
from $I_{0,A}$. 

\input{paperQDJJsub2.bbl}
%\bibliography{QDJJ,QDSN,NakajimaJJ}

\end{document}